\documentclass[a4paper,fleqn,usenatbib]{mnras}
\pdfoutput=1

\usepackage{newtxtext,newtxmath}

\usepackage[T1]{fontenc}
\usepackage{ae,aecompl}

\usepackage{graphicx}
\usepackage{amsmath}
\usepackage{amssymb}
\usepackage{commath}
\usepackage{listings}
\usepackage{placeins}
\usepackage{etoolbox}

\makeatletter
\patchcmd\@combinedblfloats{\box\@outputbox}{\unvbox\@outputbox}{}{%
  \errmessage{\noexpand\@combinedblfloats could not be patched}%
}%
\makeatother

\title[Episodic mass ejections from CE objects]{Episodic mass ejections from common-envelope objects}

\author[M. Clayton, Ph. Podsiadlowski, N. Ivanova \& S. Justham]{
Matthew Clayton$^{1}$\thanks{E-mail: matthew.clayton@physics.ox.ac.uk (MC)},
Philipp Podsiadlowski$^{1}$,
Natasha Ivanova$^{2,5}$ \&
Stephen Justham$^{3,4,5}$
\\
$^{1}$ Department of Physics, University of Oxford, Keble Rd, Oxford, OX1 3RH, United Kingdom\\
$^{2}$ Department of Physics, University of Alberta, 11322-89 Ave, Edmonton, AB, T6G2E7, Canada\\
$^{3}$ School of Astronomy \& Space Science, University of the Chinese Academy of Sciences, Beijing, China\\
$^{4}$ National Astronomical Observatories, Chinese Academy of Sciences, Beijing 100012, China\\
$^{5}$ Kavli Institute for Theoretical Physics, University of California, Santa Barbara, CA, 93106-4030, USA
}

\date{Accepted XXX. Received YYY; in original form ZZZ}

\pubyear{2017}

\begin{document}
\label{firstpage}
\pagerange{\pageref{firstpage}--\pageref{lastpage}}
\maketitle

\begin{abstract}
After the initial fast spiral-in phase experienced by a common-envelope binary, the system may enter a slow, self-regulated phase, possibly lasting 100s of years, in which all the energy released by orbital decay can be efficiently transported to the surface, where it is radiated away.  If the remaining envelope is to be removed during this phase, this removal must occur through some as-yet-undetermined mechanism. We carried out 1-d hydrodynamic simulations of a low-mass red giant undergoing a synthetic common-envelope event in such a slow spiral-in phase, using the stellar evolutionary code \texttt{MESA}. We simulated the heating of the envelope due to frictional dissipation from a binary companion's orbit in multiple configurations and investigated the response of the giant's envelope. We find that our model envelopes become dynamically unstable and develop large-amplitude pulsations, with periods in the range 3--20 years and very short growth time-scales of similar order. The shocks and associated rebounds that emerge as these pulsations grow are in some cases strong enough to dynamically eject shells of matter of up to 0.1~$\mathrm{M}_{\odot}$, $\sim 10$\,\% of the mass of the envelope, from the stellar surface at above escape velocity. These ejections are seen to repeat within a few decades, leading to a time-averaged mass-loss rate of order $10^{-3}$~$\mathrm{M}_{\odot} \: \mathrm{yr}^{-1}$ which is sufficiently high to represent a candidate mechanism for removing the entire envelope over the duration of the slow spiral-in phase.
\end{abstract}




\section{Introduction}

A common-envelope (CE) event is a phase in the evolution of a binary system wherein the binary pair orbits within an extended shared envelope \citep{Paczynski1976}. A CE configuration can arise through several processes, for example a giant star beginning dynamical mass transfer onto its companion and flooding the orbit with material, or an orbital instability in the binary such as the Darwin instability \citep{Darwin1879} causing the stars to collide. The lifetime of a CE object is limited by the action of frictional forces that dissipate energy from the binary orbit and dump it in the material of the shared envelope, causing the binary to shrink. If the binary is to survive this process, this frictional dissipation must be brought to an end by the removal of the envelope. This may be accomplished if the amount of energy transferred is large enough to unbind the envelope from the system and eject it to infinity. The binary would in such a case survive with a shortened period. Conversely, if the envelope is not ejected, the binary will continue to shrink until the two stars merge.

In either case, the CE process causes dramatic changes in the structure of a binary system and is thought to be a major formation channel for many classes of late-stage stellar systems which are of greatest interest to astrophysicists; examples include X-ray binaries, cataclysmic variables, ms pulsars, SN Ia progenitors (in both single and double degenerate models), stellar-mass gravitational-wave sources, Thorne-\.{Z}ytkow objects, and potential progenitors of both short- and long-duration gamma-ray bursts (see \citealt{Ivanova2013b} for an overview of the importance of CE events in forming stellar exotica). But CE events are by no means only a feature of unusual stellar systems; rather, they are believed to be extremely common -- it has been estimated that approximately 30 \% of all binaries undergo at least one such event during their lifetimes \citep{Han1995}.

The importance of common envelopes in the evolution of binary systems has led to a large volume of work in this area. Initial attempts to model CE events were limited to 1-dimensional studies, such as those reported by \citet{Taam1978}, \citet{Meyer1979a}, and \citet{Delgado1980}. Using such 1-d studies, three phases have been identified in the evolution of CE simulations \citep{Podsiadlowski2001, Ivanova2002}: first the loss of co-rotation phase, in which the initial embedding of the binary within the envelope occurs. This phase may occur over a long time-scale (depending on the nature of the onset of CE evolution), with the expected initial co-rotation of the system leading to low rates of energy dissipation. As co-rotation breaks down and the binary begins to shrink significantly, the system enters the rapid plunge-in phase, in which the rate of dissipation is high and the binary orbit may contract on a dynamical time-scale. The envelope may be ejected dynamically as part of this process, but if it is not, the expansion of the envelope caused by strong frictional heating reduces its density and represents a negative feedback process  controlling the rate of orbit contraction, which becomes self-regulating. This third phase -- the slow spiral-in phase -- returns to a longer time-scale as frictional dissipation in the envelope, rather than dynamical effects, determines the rate of energy dissipation. For ejection to occur in this phase, the mechanism must be dependent not on the dynamical effects of the spiral in, but rather on the thermodynamic and hydrodynamic structure of the envelope. However, the exact nature of such delayed ejections remains ill-understood. For more information on our current understanding of CE events, we recommend the recent review by \citet{Ivanova2013b}.

Despite the exponential growth in available computing power, we remain unable to model the entirety of the CE event, especially its slow phases. Although several groups of authors have performed 3-dimensional hydrodynamical simulations of CE events \citep[recent examples include][]{Ricker2012,Passy2012,Nandez2015,Ohlmann2015,Staff2016,Nandez2016,Iaconi2017}, the limitations on the physics accessible to 3-d codes, and their high computational costs, have largely restricted these simulations to the fast, dynamical plunge-in phase of CE events, in which the system evolves almost adiabatically and the evolutionary time-scales are short, although the application of 3-d studies to the beginning of the slow spiral-in phase is becoming possible \citep{Ivanova2016}. Due to these limitations, 1-d simulations remain useful tools for studying the slow, self-regulated spiral-in phase, as they are able to include more of the relevant stellar physics whilst covering the full range of time-scales involved in the CE process.

Much of the theoretical work that has been done on CE evolution has centred around attempts to construct a simple formalism for predicting the outcomes of CE events from energy considerations. However, the energy formalism, often referred to as the $\alpha$ formalism after the efficiency parameter introduced by \citet{Iben1984}, \citet{Webbink1984}, and \citet{Livio1988}, relies on having accurate expressions for all the major sources and sinks of energy present in a CE system, which are not yet fully established, and may not be possible to describe in a simple manner. In addition to sources such as orbital and thermal energy, there is strong evidence that an important part of the energy budget can be provided by the recombination energy stored in ionized material, which can be accessed by an expanding envelope to help unbind it from its star \citep[see, for example][]{Han1995,Han2002,Han2003,Ivanova2015}. Nonetheless, the details of how such liberated recombination energy may affect the ejection dynamics remain unclear \citep[see, for example, discussion in][]{Ivanova2013b}. \citet{Ivanova2013a} argued that the properties of a class of optical transients would be well explained if a large fraction of the hydrogen recombination energy in the ejected material was often radiated away rather than used to help the ejection; however the fraction of CE mass ejections to which that conclusion applies is uncertain, and their argument does not restrict the usefulness of the helium recombination energy.

In a recent paper, \citet{Ivanova2015} (hereafter I15) carried out 1-dimensional simulations of a red giant envelope during the slow spiral-in phase of a synthetic CE event. The authors of I15 were able to show the effects on a giant envelope of the heat released by an embedded binary. In many of their models the combination of this heating and the recombination energy released as the envelope expanded was able to render the envelope dynamically unstable. This result, in addition to demonstrating the importance of recombination energy to CE events, indicates that the energy input rates expected to be seen in the slow spiral-in phase are widely sufficient to destabilise the envelope and render it liable to undergo a later dynamical ejection. However, the simulations reported by I15 are based on hydrostatic stellar models, so cannot accurately describe the evolution of the envelope after it becomes dynamically unstable.

In this work, we report the results of 1-dimensional hydrodynamical simulations of a red giant primary undergoing a similar synthetic CE event, carried out with the stellar evolution code \texttt{MESA} \citep{Paxton2011, Paxton2013, Paxton2015}; we have designed these simulations to match closely the initial parameters used by I15 within this independent code. As both that study and this work make use of fully-featured stellar evolution codes, both sets of simulations contain all the relevant stellar physics which is accessible in 1 dimension, up to the fact that I15 carried out their simulations under the assumption that their giant envelope was hydrostatic. In this work, we first attempt to reproduce the hydrostatic results of I15, then we relax the hydrostatic assumption and make use of the hydrodynamics treatment available in \texttt{MESA}. This distinction allows us to follow the dynamical behaviour that arises as these models destabilise.

It is valuable to compare our approach to the study of dynamically unstable single giants, which has a long history: early papers by \citet{Lucy1967}, \citet{Roxburgh1967}, and \citet{Paczynski1968} proposed the loss of dynamical stability of giant envelopes as a trigger for the ejection of the envelope itself and the creation of a planetary nebula, and showed that such instabilities were to be expected in late-stage giants. However, these authors were imagining the ejection of the entire envelope in a single outflow event. When hydrodynamical simulations of such envelopes were performed, it was found that such direct ejections did not occur, but instead the instability was pulsational in nature \citep[see][]{Keeley1970}. Subsequent numerical studies performed by authors such as \citet{Wood1974} and \citet{Tuchman1978} found that the pulsations of such dynamically unstable giants resembled those seen in long-period variables \citep[those authors carried out their work in the context of Mira stars, see also][]{Tuchman1979} but also that they resulted in a series of repeated mass-loss events by dynamically ejecting shells of mass. More recent work has suggested that the termination of the AGB phase is due to the onset of dynamical instability and accompanying pulsations \citep[see][]{Wagenhuber1994,Han1994}. We shall compare the results of our simulations to those seen in this analogous scenario.

In Section~\ref{sec:themodel}, we describe the model star we use and the parameters of our simulations; in Section~\ref{sec:results}, we report on the results of those simulations; in Section~\ref{sec:discussion}, we discuss the interpretation and implications of these results for the study of CE evolution and the study of giant envelopes in general; and in Section~\ref{sec:conclusions}, we draw conclusions and suggest ideas for future work in this area.

\section{The Simulation Model}
\label{sec:themodel}

In 1-dimensional studies of CE events, the most commonly used technique is to co-opt a stellar evolution code to model a lone giant star, and then to simulate the presence of an embedded companion by adding quantities such as heat, gravitational mass, and angular momentum at appropriate locations within the envelope. In this manner, it is possible to study the response of the envelope to an ongoing CE event in a slow phase without being forced to make the compromises in included physics and resolution that would be necessary for a true 3-d simulation of the event. In this work, we follow I15 in adopting this method. All our calculations were performed using version 7624 of \texttt{MESA}\footnote{http://mesa.sourceforge.net} (Modules for Experiments in Stellar Astrophysics), a state-of-the-art open-source stellar evolution code \citep{Paxton2011, Paxton2013, Paxton2015}.

The initial model we use for our simulations has been chosen to match as closely as possible the one used in I15 -- a 1.6~$\mathrm{M}_{\odot}$ red giant. As that work used a different stellar evolution code, small differences in the model cannot be avoided. We chose to match our giant's core mass to the value used in I15, and tuned the mixing length parameter $\alpha$ to obtain approximately the 100~$\mathrm{R}_{\odot}$ star used in that paper, using the Cox \& Giuli formulation of mixing length theory \citep[see][]{Cox1968}. We evolved a solar metallicity star of 1.6~$\mathrm{M}_{\odot}$ from the zero-age main sequence through to the red giant branch, stopping the star's evolution when its helium core, which we define as that region within which the fractional density of hydrogen $X$ drops below $10^{-10}$, reached 0.422~$\mathrm{M}_{\odot}$. Selecting an $\alpha$ value of 1.95 gave this star a radius of 100.7 R. We adopted a simple Eddington grey atmosphere model and neglected wind mass loss. We shall refer to this giant model as the ``initial model'', which serves as the starting point for all simulations reported below.

In order to study the response of this initial model to a simulated CE event, we must emulate the presence of an embedded binary companion within the envelope. This is accomplished by injecting additional heat into the envelope to represent the frictional dissipation of the binary's orbital energy during a self-regulated spiral-in. The distribution of such heating during CE evolution is not well understood \citep[see][]{Taam2000}, with multiple processes involved which are expected to dump heat in different regions of the envelope. For example, viscous drag forces acting on the secondary as it moves through the material of the envelope will tend to dissipate heat at or near the radius of the companion's orbit \citep[as in][]{Taam1978}, whereas viscous shear between layers of a differentially rotating envelope will tend to cause heating in an extended region outside the orbit \citep[as in][]{Meyer1979a}. We again follow I15 in adopting two alternate prescriptions for the heating. In one case, which we refer to as ``base heating'', heat is added to a thin layer at the base of the envelope, where an embedded companion is expected to orbit during a self-regulated spiral in. In contrast, we also have the ``uniform heating'' case, in which heat is added throughout the entire convective envelope of the star. These two prescriptions therefore represent the opposite extremes of heat distribution during a CE event. In both cases, heat is added at a constant rate per unit time.

In both of our heating cases, we define the heating region by Lagrangian mass coordinate within the star (where we define the mass coordinate of a shell within the star as being equal to the total amount of mass contained within that shell). These layers are fixed at specific mass coordinates at the beginning of our simulations and move with the Lagrangian motion of the stellar material. Heat is then added within the specified region at a uniform rate per unit mass. The base heating case injects heat into a layer 0.1~$\mathrm{M}_{\odot}$ thick, starting at the base of the initial convective envelope at a mass coordinate of 0.4257~$\mathrm{M}_{\odot}$, and ending at 0.5257~$\mathrm{M}_{\odot}$, whereas the uniform heating case has the same inner boundary but has its outer edge at 1.5995~$\mathrm{M}_{\odot}$, at the top of the convective zone. Rather than heating beginning sharply at the edges of these regions, both cases smooth the rate of heat injection linearly over a region 0.01~$\mathrm{M}_{\odot}$ in mass on the inside of their boundaries.

The rates at which we have added heat to the envelope replicate those used in I15, which are intended to simulate the slow spiral-in of a 0.3~$\mathrm{M}_{\odot}$ companion. If the slow spiral-in phase begins when the secondary is located at the base of the giant's convective envelope, the orbit will possess approximately $10^{48}$~ergs of energy. A reasonable range of time-scales for the deposition of this energy into the envelope is 10--1000 years, which gives us energy deposition rates of between $10^{45}$ and $10^{47}$~$\mathrm{ergs \: yr}^{-1}$. This added heat and the star's existing nuclear heating (which is almost completely constant on the time-scales we're dealing with) are the only energy sources present in our model.

In addition to the two heating cases, the simulations we present are run either with or without the use of a hydrodynamics scheme. In cases without a hydrodynamic treatment, the inertial terms in the stellar pressure equation are ignored, which amounts to an assumption that the star is in hydrostatic equilibrium. In our hydrodynamical simulations, we make use of \texttt{MESA}'s explicit hydrodynamics treatment. \texttt{MESA} also has an implicit hydrodynamics implementation, which is designed to improve the accuracy of energy conservation; however, we found this implementation to be unsuitable for our purposes due to convergence difficulties. In all our hydrodynamic simulations, we apply artificial viscosity in order to allow \texttt{MESA} to resolve hydrodynamic shocks; we adopt a shock width of 5\% of the local radius for this purpose \citep[$l_2$ in section~4.2 of ][]{Paxton2015}. Variations of the numerical parameters of our simulations were found not to affect the qualitative nature of our results.

In our hydrodynamic simulations, unless otherwise stated, a custom mass-loss routine is used to remove unbound material which is expanding at above the local escape velocity: whenever a contiguous layer develops at the surface of the model containing matter which exceeds the local escape velocity at every point, the material of that layer is removed from the model's surface. We achieve this in \texttt{MESA} by implementing a wind-loss scheme which removes this unbound matter from the surface exponentially with a time constant of 0.01 years. At each timestep, we find the mass $m$ of the ejecting layer (if one is present), and apply mass loss at a rate of $100m$ per year. The algorithm used to generate this mass loss is available in Appendix~\ref{app:mdot} This mass-loss rate was chosen to ensure that the time-scale for mass removal is at least an order of magnitude below the dynamical and thermal time-scales of the star at all times, whilst not being so fast as to introduce artefacts into the dynamics of the star. In practice, the removal of an entire escaping layer occurs over approximately a month of star time.

If mass within the heating zone is removed by our mass-loss routine, the specific heating rate increases to keep the total heating rate constant.

\section{Simulation Results}
\label{sec:results}

\subsection{Hydrostatic simulations}

\begin{figure}
	\includegraphics[width=\columnwidth]{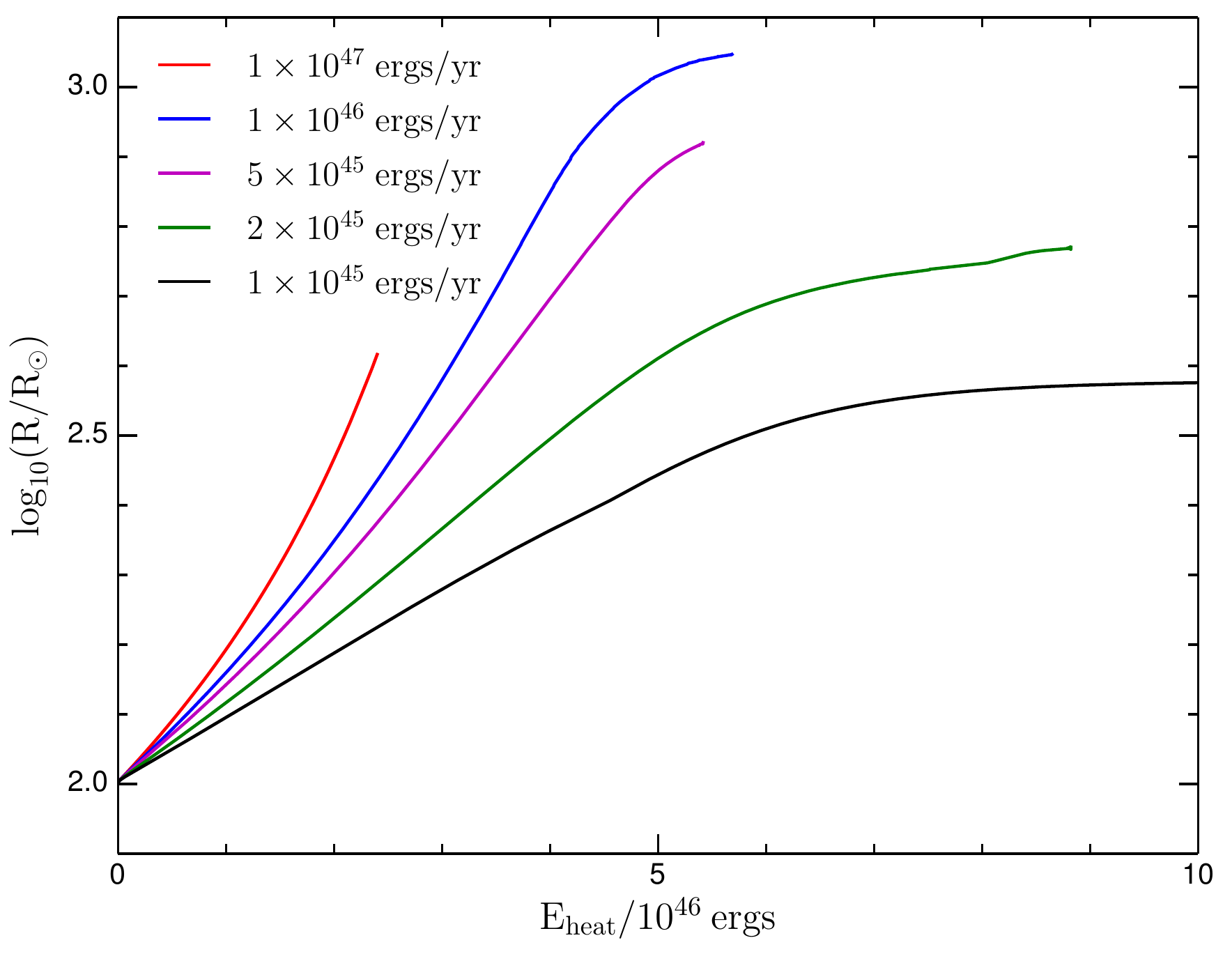}
    \caption{The radii of hydrostatic models as a function of the total heating energy deposited where the artificial heating is uniformly spread throughout the convective envelope. The blue, pink and green cases terminate due to convergence failures, the red case is terminated when the surface velocity reaches twice the escape velocity.}
    \label{fig:fig1}
\end{figure}

\begin{figure}
	\includegraphics[width=\columnwidth]{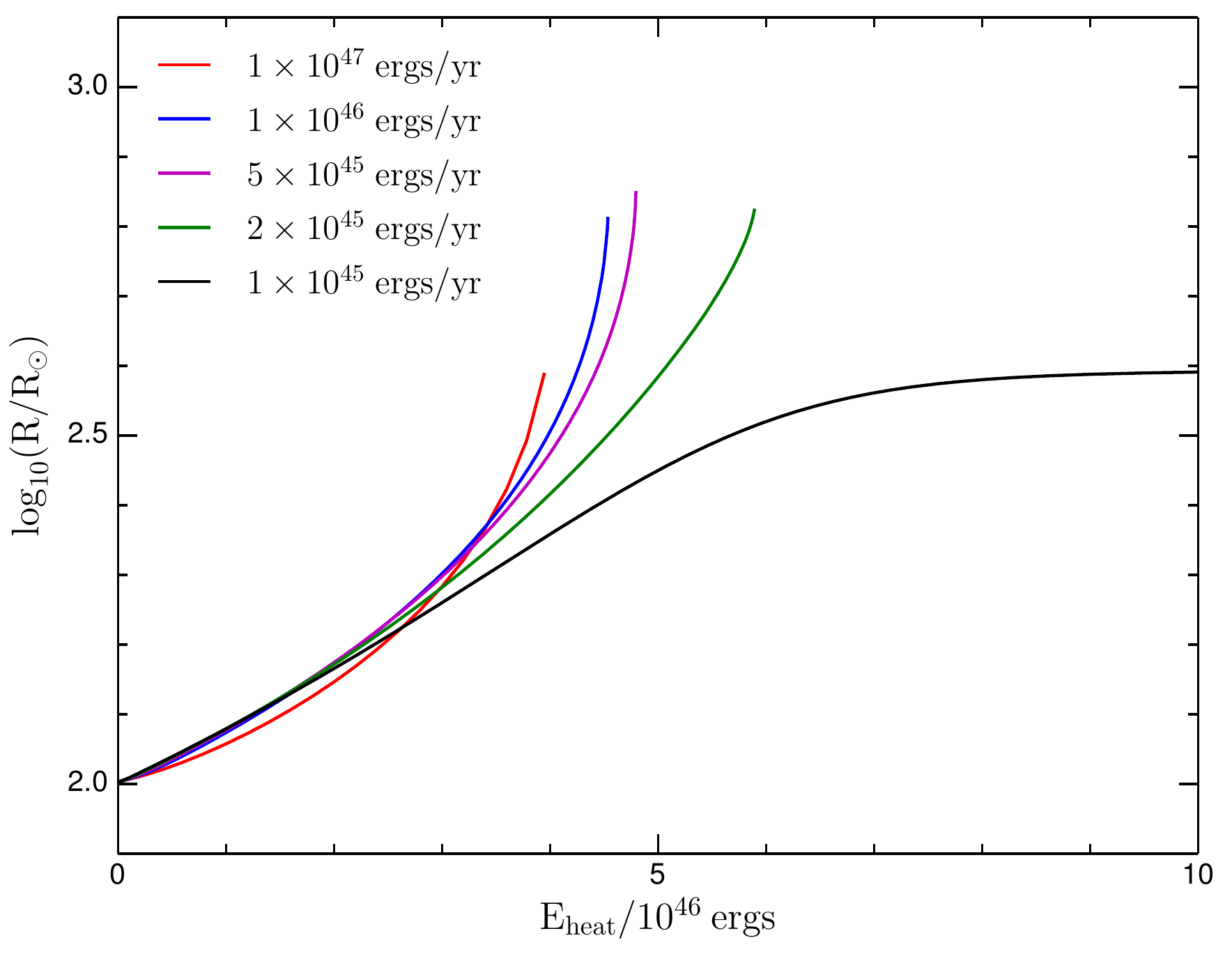}
    \caption{The same as Fig.~\ref{fig:fig1} for the case where the artificial heating is restricted to the base of the convective envelope. Simulations are again terminated when the surface velocities reach twice the escape velocity.}
    \label{fig:fig2}
\end{figure}

We carried out hydrostatic simulations in both uniform heating and base heating cases with heating rates of $10^{45}$, $2 \times 10^{45}$, $5 \times 10^{45}$, $10^{46}$, and $10^{47}$~$\mathrm{ergs \: yr}^{-1}$. The evolution of the outer radii of these models can be seen in Figs~\ref{fig:fig1} and \ref{fig:fig2}. These simulations agree closely with the results presented in I15, despite the independent evolution code. A thorough analysis of the physics seen in these simulations can be found in that paper, we will merely perform a brief comparison.

As in I15, we see that in the base heating case only the lowest energy deposition rate leads to a stable model, with all higher heating rates causing an accelerating expansion of the star that quickly reaches and exceeds escape velocity (these simulations were all terminated when the surface reached double the escape velocity). Of course, this outcome is only meaningful in a hydrostatic simulation insofar as it shows us that the hydrostatic assumption is inappropriate in this regime due to instability of the envelope, and we must switch to a hydrodynamic treatment.

The uniform heating cases also tend to match the results of I15, but with one notable difference: in I15, the models with the second and fourth highest heating rates, $10^{46}$ and $2 \times 10^{45}$~$\mathrm{ergs \: yr}^{-1}$, expanded into stable equilibria; however, the model in between these two, $5 \times 10^{45}$~$\mathrm{ergs \: yr}^{-1}$, also expanded to its equilibrium luminosity but then destabilised and failed to converge -- an outcome expected when there exists no unique hydrostatic and thermal equilibrium solution. As this numerical instability is thought to reflect a real, physical instability in the envelope, the greater numerical stability of the model with the faster heating rate is unexpected. We do not recover this result; instead, in our simulations the $10^{46}$ and $5 \times 10^{45}$~$\mathrm{ergs \: yr}^{-1}$ models both destabilise, with the larger heating rates destabilising faster (although the $5 \times 10^{45}$~$\mathrm{ergs \: yr}^{-1}$ case appears to destabilise first in Fig.~\ref{fig:fig1}, as the \emph{x}-axis of that figure is in energy, the higher heating rate does destabilise first in time). This outcome is more aligned with the natural expectation that greater energy deposition rates are more destabilising to the model. The lines in Fig.~\ref{fig:fig1} representing these three models all end where the simulations terminate due to convergence failures; by lengthening the timesteps used in the simulations it is possible to artificially delay these failures somewhat, or in the $5 \times 10^{45}$~$\mathrm{ergs \: yr}^{-1}$ case, to suppress the failure entirely, in which case the models remain for a while in unstable equilibrium. The simulation with the very largest rate of uniform heating experiences the same rapid expansion as seen in the equivalent base heating case and quickly exceeds double the escape velocity, whereupon that simulation is terminated.

The runaway expansion seen in both uniform and base heating cases and the destabilising of the higher uniform heating rates all clearly indicate that the models in question cannot be hydrostatically stable, and that we must use a hydrodynamic treatment instead to perform meaningful simulations of the evolution of these stars after they destabilise. In fact, as we shall see shortly, the use of a hydrodynamic code leads to significant differences in the behaviour of the majority of these models, even before the point where their behaviour becomes obviously dynamical.

\subsection{Hydrodynamic simulations}
\label{sec:hydrosims}

\begin{figure*}
	\includegraphics[width=1.6\columnwidth]{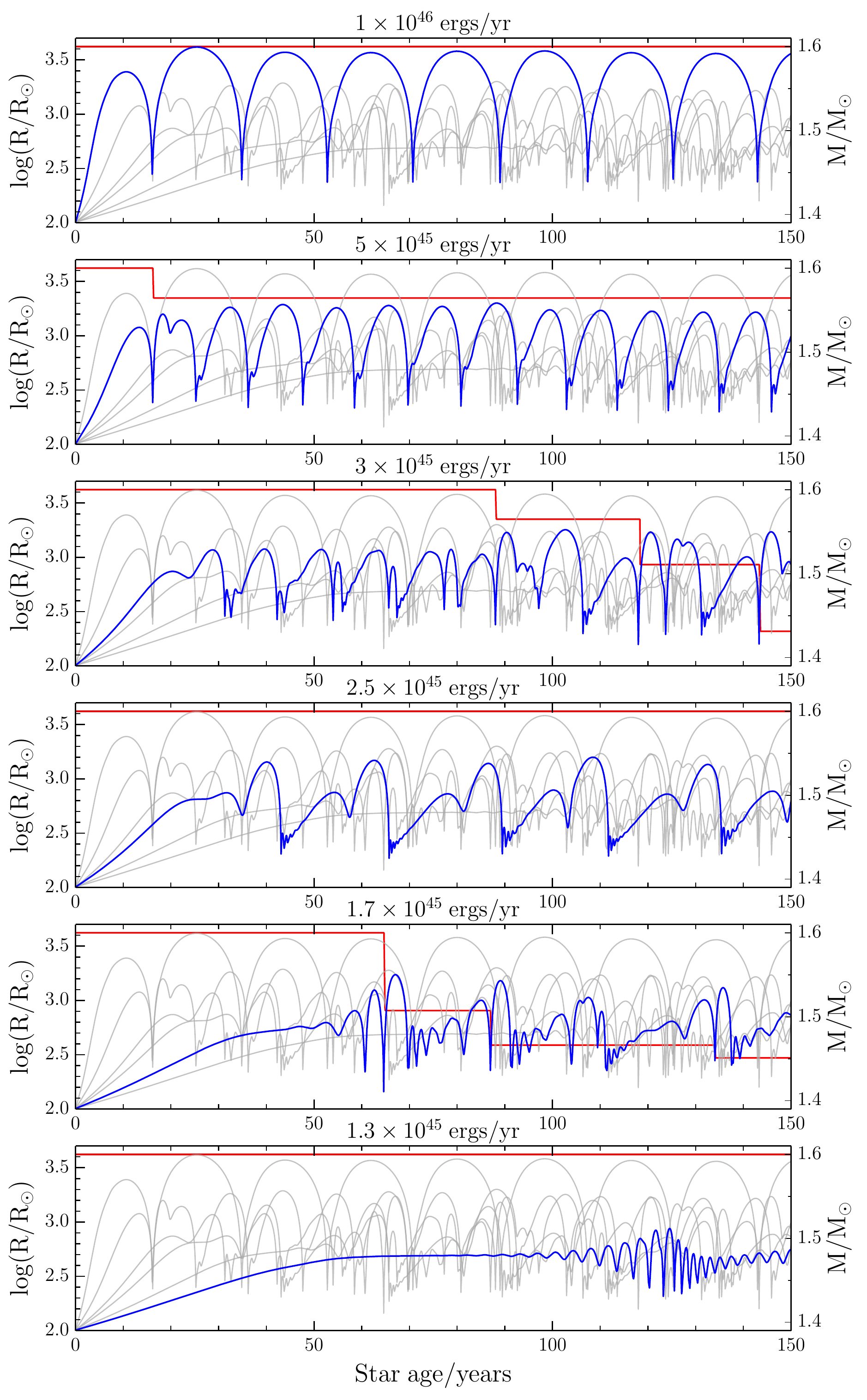}
    \caption{The surface radii (blue) and masses (red) of a selection of hydrodynamical models in the uniform heating case, chosen to be representative of the behaviours observed in these simulations. Each subplot describes a model heated at a different rate. The relevant heating rate appears above each subplot. Sudden mass changes accompany dynamical ejection events, where the outer layers of the model have exceeded escape velocity and have been removed (see Section~\ref{sec:hydrosims}). The grey lines are the radius evolutions of the other heating rates, overplotted for comparison.}
    \label{fig:fig3}
\end{figure*}

\begin{figure*}
	\includegraphics[width=1.6\columnwidth]{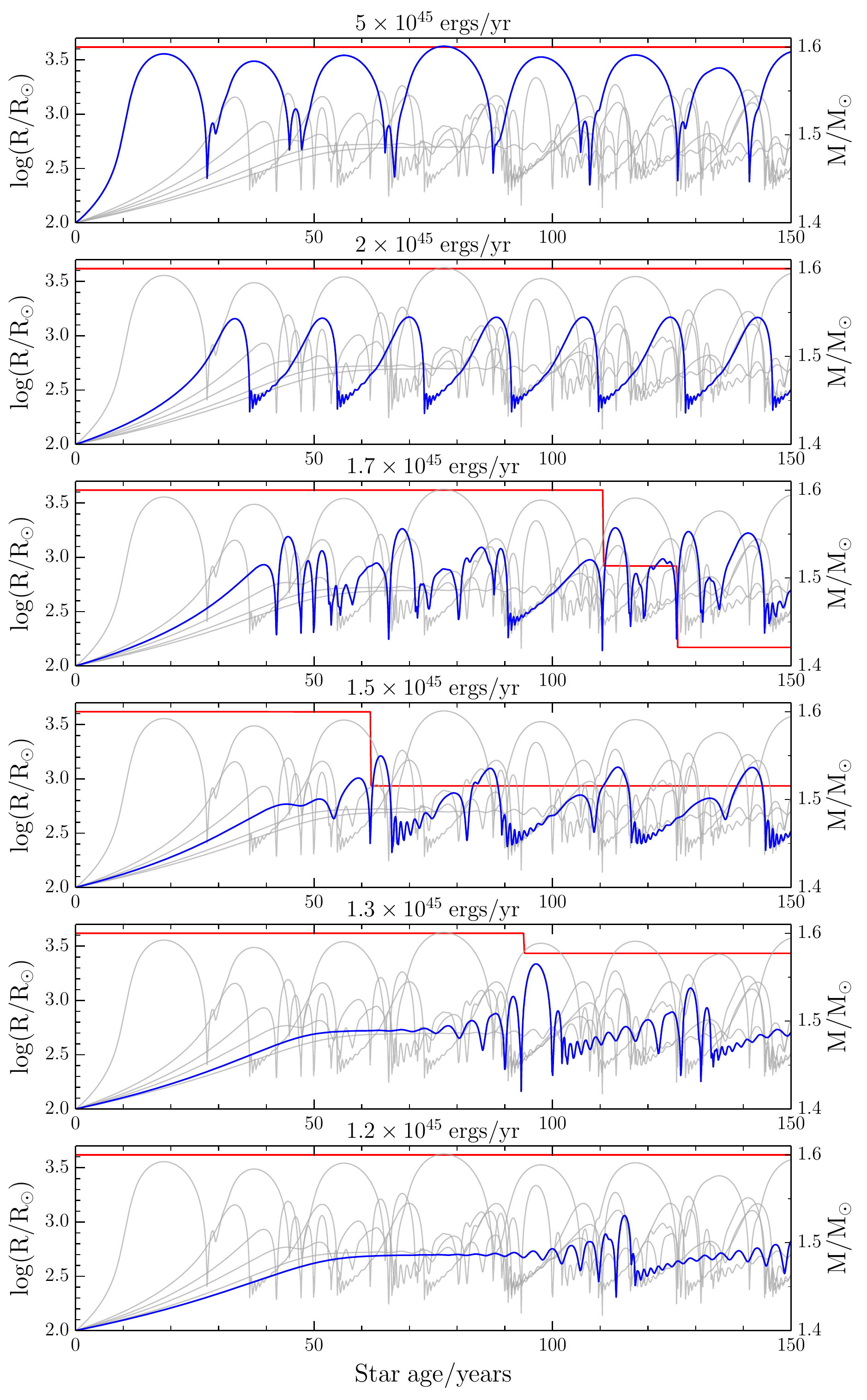}
    \caption{The surface radii (blue) and masses (red) of a selection of hydrodynamical models in the base heating case, chosen to be representative of the behaviours observed in these simulations. Each subplot describes a model heated at a different rate. The relevant heating rate appears above each subplot. Sudden mass changes accompany dynamical ejection events, where the outer layers of the model have exceeded escape velocity and have been removed (see Section~\ref{sec:hydrosims}). The grey lines are the radius evolutions of the other heating rates, overplotted for comparison.}
    \label{fig:fig4}
\end{figure*}

The hydrostatic models reported on above make it clear that a hydrodynamic treatment is necessary to understand the evolution of these unstable models. We carried out a large number of hydrodynamical simulations in both the base and uniform heating cases with different heating rates. A selection of these simulations showing the regimes of behaviour that arise are presented in Figs~\ref{fig:fig3} \& \ref{fig:fig4}, which show the evolution of the radius and mass of our simulated giant over time for different heating rates. The very lowest and highest heating rates shown in our hydrostatic simulations do not appear in these figures, as they behave the same as the hydrostatic cases: an expanded equilibrium for a heating rate of $10^{45}$~$\mathrm{ergs \: yr}^{-1}$, and a direct ejection in the case of $10^{47}$~$\mathrm{ergs \: yr}^{-1}$. For intermediate heating rates, however, occupying almost the entire parameter space between these extremes, we see the emergence of large scale pulsations which come to dominate the evolution of these models.

The behaviour of these pulsations varies strongly with heating rate, and the cases shown in Figs~\ref{fig:fig3} \& \ref{fig:fig4} are a sample chosen to demonstrate the different regimes that emerge. In all cases in which pulsations develop (which is all heating rates other than some very low values just above our minimum and some rapidly ejecting models just below the maximum), we see the rapid, exponential growth of those pulsations, quickly reaching very large amplitudes and becoming supersonic. These pulsations continue to grow until damped by non-linear effects (discussed in section~\ref{sec:phases} below). The growth rate of the pulsations increases with heating rate and can reach extremely high values, with the growth time-scale reaching the order of the pulsation period for higher heating rates and the amplitudes attaining limiting size within one pulsation cycle.

The behaviour of the pulsations seen in each our models falls into one of three regimes:

\begin{enumerate}
\setlength\itemsep{1em}
\item
\textbf{Self-limiting --} For low heating rates, such as $1.3\times10^{45}$~$\mathrm{ergs \: yr}^{-1}$ in the uniform heating case (Fig.~\ref{fig:fig3}) and $1.2\times10^{45}$~$\mathrm{ergs \: yr}^{-1}$ in the base heating case (Fig.~\ref{fig:fig4}), the pulsations grow exponentially over multiple pulsation cycles until non-linear effects act to limit their amplitude (usually at the point at which these pulsations become supersonic and shocks begin to develop during the compression phases of the pulsations). These pulsations then proceed to grow again in cycles of alternating growth and modulation.

\item
\textbf{Ejecting --} For intermediate heating rates, such as $1.7\times10^{45}$~$\mathrm{ergs \: yr}^{-1}$ for uniform heating and $1.5\times10^{45}$~$\mathrm{ergs \: yr}^{-1}$ for base heating, the pulsation amplitudes grow sufficiently large and the velocities sufficiently supersonic that the compressions and accompanying shocks which occur are strong enough to dynamically eject shells of matter from the surface of the models at greater than the star's escape velocity. These ejected shells, which can be up to $\sim0.1 \; \mathrm{M}_{\odot}$, are completely unbound and represent a form of dynamical mass loss from the star. The matter in these ejected shells is excised from the simulation after it exceeds escape velocity, and we follow the evolution of the matter which remains bound. The points at which mass ejections occur can therefore be seen by the rapid changes in remaining model mass (the red line) in Figs~\ref{fig:fig3} \& \ref{fig:fig4}, where a shell of material has been removed. The ejection phenomenon is discussed in more detail in Section~\ref{sec:ejections} below.

\item
\textbf{Non-ejecting --} In this third regime, seen in cases such as $2.5\times10^{45}$~$\mathrm{ergs \: yr}^{-1}$ and $10^{46}$~$\mathrm{ergs \: yr}^{-1}$ for uniform heating, and $2\times10^{45}$~$\mathrm{ergs \: yr}^{-1}$ for base heating, pulsations reach amplitudes large enough to launch ejections, yet no ejections occur. Instead, the shocks which develop in these cases dissipate the energy of the pulsation and excite higher-order pulsations which then decay away. The primary pulsation proceeds to grow once again, resulting in stable repeating cycles. These non-ejecting cycles can be seen for the largest heating rates, but also in cases such as $2.5\times10^{45}$~$\mathrm{ergs \: yr}^{-1}$ with uniform heating, when both lower and higher heating rates are seen to exhibit ejections. It is also possible for a simulation to exhibit one or more ejections before settling into a repeating non-ejecting cycle. This phenomenon, and why the switch between ejecting and non-ejecting behaviours is not a simple bifurcation, is discussed in Section~\ref{sec:tobe} below.

\end{enumerate}

\begin{figure*}
	\includegraphics[width=2.1\columnwidth]{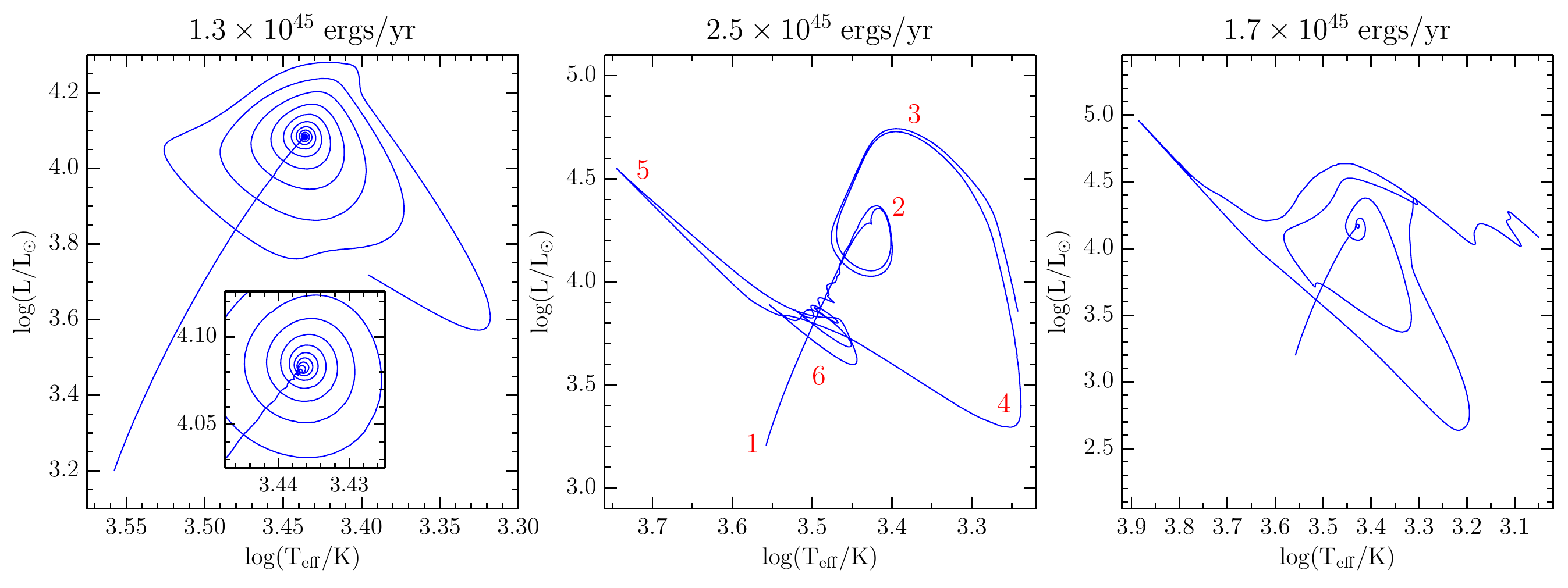}
    \caption{Three example Hertzsprung-Russell tracks for uniform heating cases, all of which appear in Fig.~\ref{fig:fig3}. Left: a case with no ejections, plotted until maximum amplitude is attained, Centre: A case that exhibits a repeated cycle of strong envelope shocks but no ejections, plotted until after the first complete cycle. This track begins at point 1 and progresses in numerical order to point 6, whereupon it moves back to point 2 and repeats. Right: A case that undergoes a large ejection, with our mass-loss scheme being deactivated so the ejecting shell's surface is seen.}
    \label{fig:fig5}
\end{figure*}

In order to understand the physics behind the pulsations we observe, it is instructive to examine the shape of the pulsations in the Hertzsprung-Russell (HR) diagram, examples of which are shown in Fig.~\ref{fig:fig5}. Three simulations with uniform heating at different rates can be seen in this figure (all of these simulations also appear in Fig.~\ref{fig:fig3}). On the left, we have relatively slow pulsation growth for the low heating rate of $1.3 \times 10^{45}$~$\mathrm{ergs \: yr}^{-1}$ up to the point where the pulsation reaches its maximum amplitude. These pulsations produce circles in the HR diagram initially, but as the amplitude increases we can see the emergence of the characteristic shape seen in the other two HR diagrams as the pulsation enters the non-linear regime. A discussion of the non-linear effects which appear in large amplitude pulsations can be found in Section~\ref{sec:phases}.

The central plot of Fig.~\ref{fig:fig5} shows the $2.5 \times 10^{45}$~$\mathrm{ergs \: yr}^{-1}$ case, also for uniform heating, which does \emph{not} exhibit ejections despite the large amplitudes attained and the emergence of strong shocks in the envelope. One complete sequence of the growth and dissipation of the primary pulsation is shown. The pulsation growth of this model is extremely rapid, reaching limiting amplitude in less than two complete cycles. The initial expansion of the model from its unperturbed state, points 1--2 in the plot, is followed by  pulsation around the ``equilibrium'' point labelled 2. The expansion phase at maximum amplitude coincides with point 3; this expansion phase is very long due to the increase in pulsation period that occurs with increasing radius. The giant envelope then suffers a cooling catastrophe (see below), leading to the low temperatures seen at point 4. When contraction occurs at this amplitude, the outer layers contract at supersonic speeds, and a shock is formed as the pulsation approaches minimum radius. This shock moves outward through the envelope and breaks out from the surface, producing the dramatic but short-lived spike in temperature and luminosity at point 5. Due to the radiative losses sustained during the expansion phases, the model left after the shock is not in thermal equilibrium, so it then proceeds back up its original expansion trajectory from points 6--2, after which the pulsation begins to grow again. The higher-order pulsations excited by the shock can be seen around point 6, decaying as the model expands once again.

The right hand plot in Fig.~\ref{fig:fig5} shows the $1.7 \times 10^{45}$~$\mathrm{ergs \: yr}^{-1}$ case for uniform heating, which \emph{does} dynamically eject material. To generate this plot, our mass-loss routine was deactivated, so the ejected layer continues to be simulated as it expands away. The same features can be seen as in the non-ejecting case next to it, but with the luminosity peak at shock breakout followed by an ejection. As the HR diagram only tracks properties of the model's surface, cases when that surface is ejected to infinity trail off into the cold region of the diagram. Details of the properties of this surface (the ``wiggles'' seen in the plot) depend on the model's outer boundary conditions, which become increasingly less applicable in the unbound regime, so are not expected to be meaningful.

\subsection{Non-linear physics in the pulsation cycle}
\label{sec:phases}
In our analysis, we will make use of the envelope's dynamical time-scale, for which we shall use the formula
\begin{equation}
\tau_\mathrm{dyn} \approx \sqrt{\frac{R_\star^3}{G M_\star}},
\end{equation}
and its radiative cooling time-scale, for which we use
\begin{equation}
\tau_\mathrm{KH} \approx \frac{G M_\star M_\mathrm{env}}{2 R_\star L_\star},
\end{equation}
where $R_\star$ is the radius of the star, $G$ is Newton's gravitational constant, $M_\star$ is the total mass of the star, $M_\mathrm{env}$ is the mass of the envelope, and $L_\star$ is the luminosity of the star. Whenever references are made to these time-scales, the values of $R_\star$, $L_\star$, $M_\mathrm{env}$ and $M_\star$ which are used are instantaneous, rather than initial values.

\begin{figure}
	\includegraphics[width=1.05\columnwidth]{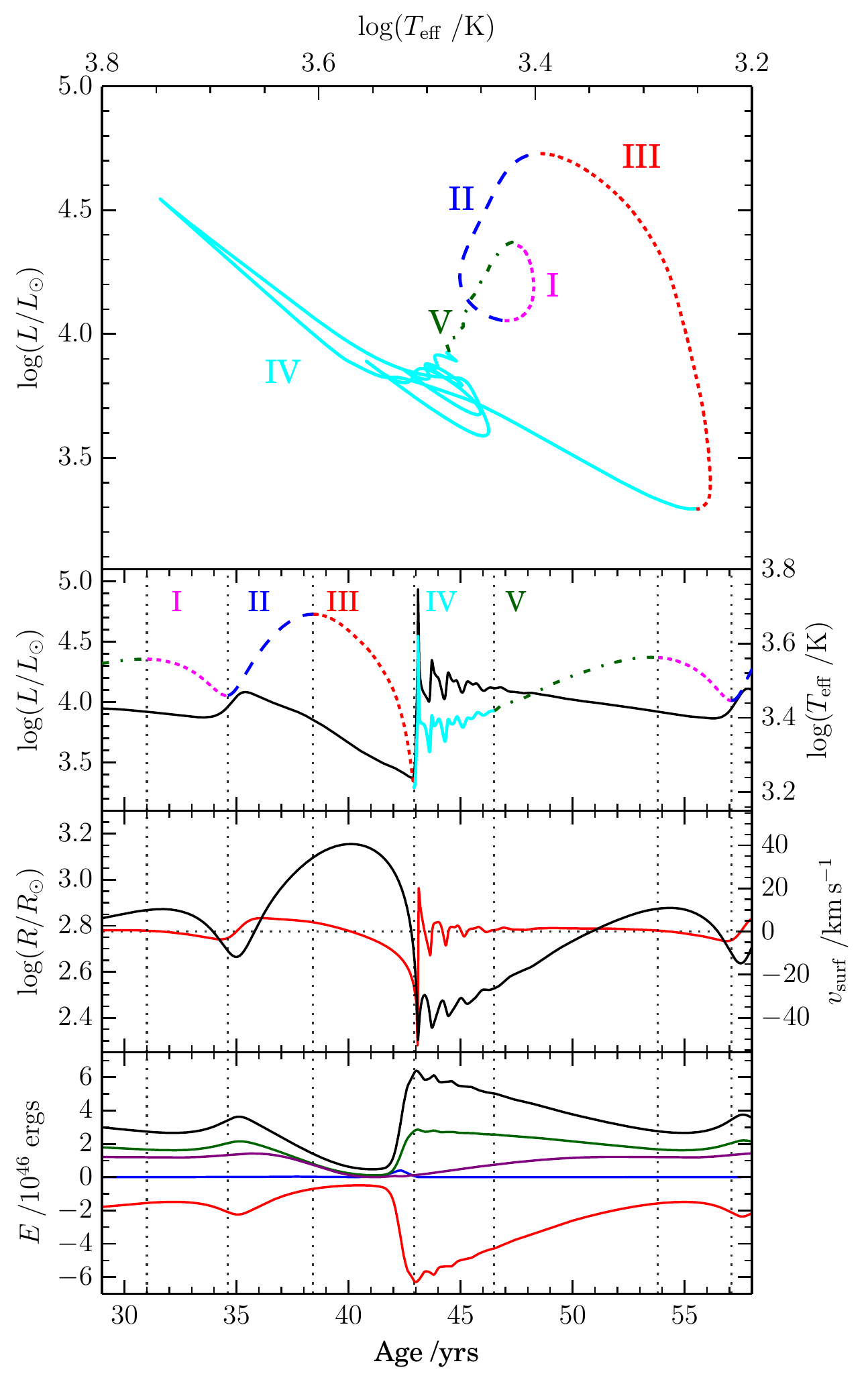}
    \caption{The pulsation phases dominated by non-linear effects in the $2.5 \times 10^{45}$~$\mathrm{ergs \: yr}^{-1}$ uniform heating model. The first pulsation cycle experienced by this model is shown on a Hertzsprung-Russell diagram (top); we show the evolution of the model's luminosity in variable colours and effective temperature in black (second from top); below this is shown the model's radius in black and surface velocity in red (third from top); and in the bottom plot we show the internal energy (thermal + recombination) in black, ionization energy in green, kinetic energy in blue, gravitational energy in red, and total energy in purple,  summed over the giant envelope between a mass coordinate of 0.43~$\mathrm{M}_\odot$ and the surface.}
    \label{fig:fig6}
\end{figure}

As the pulsation amplitudes in our simulations grow, they become dominated by non-linear effects. To investigate how important these effects are, we will examine the repeating cycle appearing in the central simulation plotted in Fig.~\ref{fig:fig5} in more detail. This simulation ($2.5 \times 10^{45}$~$\mathrm{ergs \: yr}^{-1}$ uniform heating) is shown in Fig.~\ref{fig:fig6}, which plots the evolution of the model's surface properties during this cycle, as well as the values of the internal, gravitational, ionization and kinetic energy of the envelope. In this figure, the pulsation cycle has been split into 5 phases:

\begin{enumerate}
\item \textbf{Phase I -- Initial compression --} The amplitude in this phase is too low for the model to experience significant non-linear effects; the amplitude grows exponentially in time.

\item \textbf{Phase II -- Expansion --} In this phase the amplitude of the pulsation continues to increase and nears its maximum. The pulsation period lengthens as the model's radius grows, because the dynamical (freefall) time-scale of the envelope increases with radius. The increase in surface area leads to a reduction of the envelope's radiative cooling time-scale at the same time. As the star's radius approaches its maximum, the radiative cooling time-scale of the envelope actually drops below its dynamical time-scale -- for example, at the end of phase II in Fig.~\ref{fig:fig6}, the dynamical time-scale is 646 days, but the radiative cooling time-scale is only 161 days. The envelope can therefore approach thermal equilibrium (cool exponentially towards an equilibrium temperature set by its radius) faster than it can approach hydrostatic equilibrium (collapse to an equilibrium radius set by its thermal properties).

The increase in pulsation period with radius also causes the pulsation to begin to decohere, as the outer layers of the envelope, being at a larger radius, pulsate with a longer period than layers deeper within the star. This causes the inner layers of the envelope to pulsate out of phase with the outer layers,  with the inner layers reaching their maximum radii and beginning to contract before the layers outside them do.

\item \textbf{Phase III -- Cooling catastrophe --} During this phase, the results of the envelope's short cooling time-scale become manifest. The envelope effectively cools faster than it contracts, leading to an exponential decrease in temperature down to extremely low values, and as a result of this cooling, there is a near-complete loss of pressure support to the outer layers. This leads to an extremely fast collapse of matter which is practically in freefall, whilst having lost most of its internal energy. During this phase, almost the entire envelope ($\gtrsim 1.1 \mathrm{M}_\odot$) is fully neutral, having radiated away both its thermal energy and its ionization energy (this can be seen in the bottom plot of Fig.~\ref{fig:fig6}, where the ionization energy (green) drops almost to zero in this phase). This process constitutes a cooling catastrophe, with energy transport from the inner layers of the star unable to prevent the cooling of the outer envelope to very low temperatures.

During this contraction phase, the pulsation continues to decohere, with internal layers reaching minimum radius whilst material further out is still collapsing.  The phase lag between different layers of the star leads to the development of large relative velocities and to the formation of shocks.

\item \textbf{Phase IV -- Shock breakout and ringdown --} As the envelope's collapse approaches its minimum radius, its outer layers are highly supersonic, and a strong compression shock is formed, which moves outwards towards the surface as rapidly infalling layers of the envelope collide with the material interior to them which has already decelerated. The shock is effective at converting the kinetic energy of the infalling material into internal energy, helping to reheat and reionize material in the envelope which lost almost all of its internal energy during the cooling catastrophe. The shock travels towards the surface over approximately a year, as the decoherence of the different layers within the envelope leads to the inner layers reaching minimum radius long before the outer layers (in this case, approximately one year before). As the shock reaches the surface, it produces a dramatic but short-lived spike in the luminosity and temperature of the model, and excites higher-order, shorter-period pulsations: the presence of the shock near the envelope's surface leads to the deposition of the pulsation's kinetic energy less deep within the envelope than would have occurred if the contraction had been coherent (that is, if all layers of the envelope had been in phase and reached their minimum radii at the same time). This deposition of energy excites higher-order pulsation modes, which are not effectively driven and therefore decay as the envelope rings down.

\item \textbf{Phase V -- Quasistatic reexpansion --} The large amount of radiative energy lost during phases II \& III causes the star to regain hydrostatic equilibrium at a lower energy state than the unstable state about which the observed pulsations can grow. The star therefore expands quasistatically towards this unstable state before pulsations can begin to grow again. This expansion is analogous to the initial expansion of the model from its starting state.

\end{enumerate}

\begin{figure}
	\includegraphics[width=1.05\columnwidth]{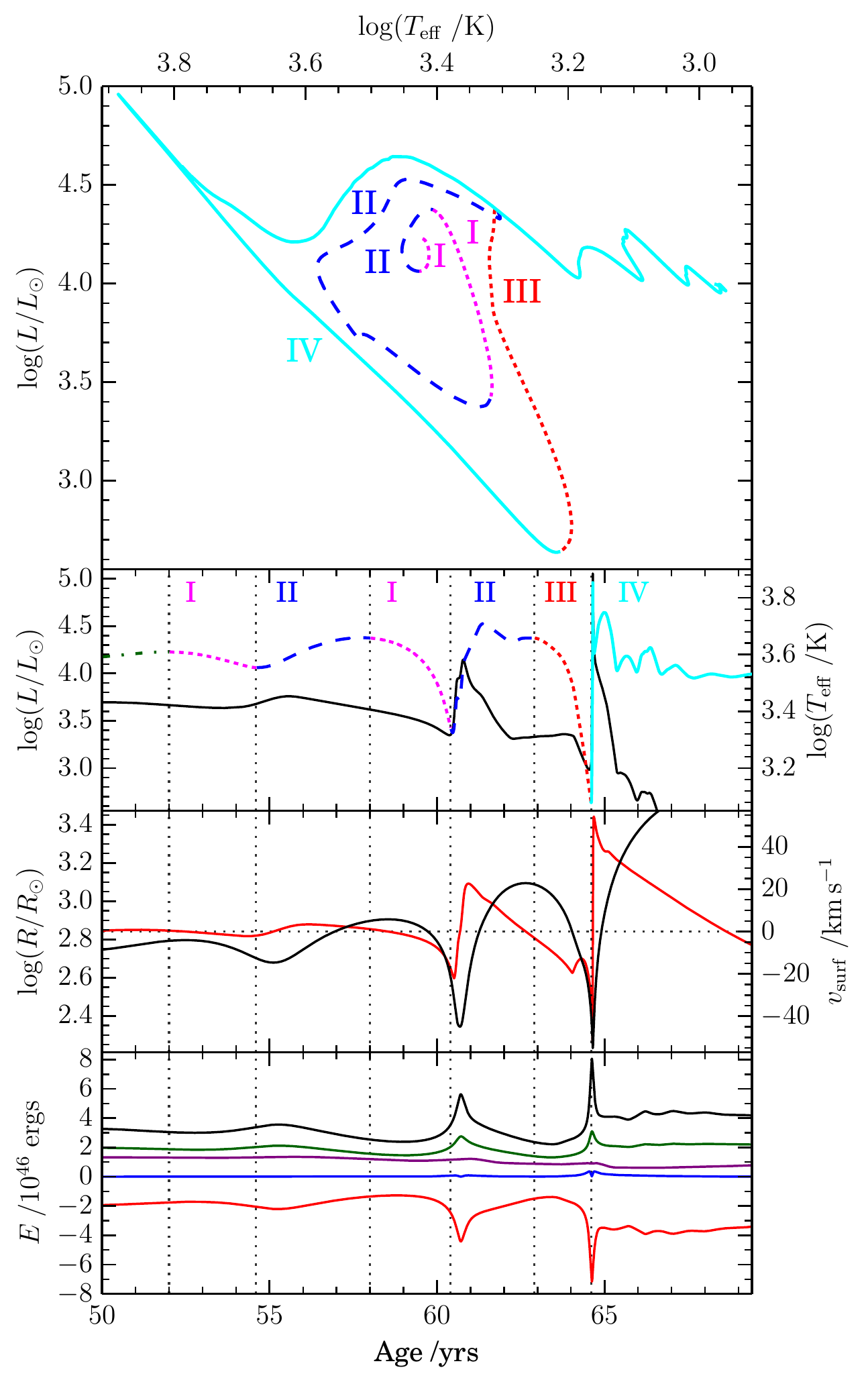}
    \caption{The pulsation phases that appear in the (ejecting) $1.7 \times 10^{45}$~$\mathrm{ergs \: yr}^{-1}$ uniform heating model. This ejecting model has our custom mass-loss routine turned off, so that after the ejection is launched the unbound mass remains part of the model, and the surface properties refer to the outer edge of this material. The first dynamical ejection experienced by this model is shown on a Hertzsprung-Russell diagram (top); we show the evolution of the model's luminosity in variable colours and effective temperature in black (second from top); below this is shown the model's radius in black and surface velocity in red (third from top); and in the bottom plot we show the internal energy (thermal + recombination) in black, ionization energy in green, kinetic energy in blue, gravitational energy in red, and total energy in purple,  summed over the giant envelope between a mass coordinate of 0.43~$\mathrm{M}_\odot$ and the surface. Note that as this diagram covers two pulsation periods, phases I and II appear twice.}
    \label{fig:fig7}
\end{figure}

The simulation used for this analysis is an example of a non-ejecting case. In the cases which do display ejections, the expansion phase typically reaches a lower maximum radius and lasts for less time. This has two effects: it prevents the internal layers of the star from decohering to as great an extent, which leads to the compression shock being stronger near the surface; and it gives the star less time to radiate away energy during the cooling catastrophe. This leads to the star retaining sufficient energy to rebound quickly and launch an ejection, rather than having all pulsation energy used up in reheating cool envelope material. Fig.~\ref{fig:fig7} shows an equivalent analysis of pulsation phases for a simulation undergoing a mass ejection (specifically, it shows the first ejection experienced by the $1.7 \times 10^{45}$~$\mathrm{ergs \: yr}^{-1}$ uniform heating model), in which the shorter expansion phase can be seen to be accompanied with much higher total and ionization energies in the envelope during that phase.

\subsection{Dynamical ejections}
\label{sec:ejections}

To gain greater insight into the physics of the ejection-launching process, we will examine a specific example in more detail. A shell ejection experienced by the $1.7 \times 10^{45}$~$\mathrm{ergs \: yr}^{-1}$ uniform heating model is shown in Figs~\ref{fig:fig10}, \ref{fig:fig11} \& \ref{fig:fig12}. This simulation was performed with our mass-loss scheme turned off so that the ejected material  appears in the plots. These plots record the values of 12 output variables throughout the giant envelope during the ejection process. In the case of variables whose range of interest spans several orders of magnitude and can be both positive and negative, we have made use of the log modulus transformation as described by \citet{John1980}, which we will refer to as the logmod function:
\begin {equation}
\label{equ:logmod}
\mathrm{logmod}(x) = \mathrm{sgn}(x)\,\mathrm{log}_{10}(\abs{x} + 1),
\end {equation}
which is linear for small values and logarithmic for large values of $x$, whilst maintaining sign and symmetry about the zero point.

We also define the pressure-weighted, volume-averaged value of the first adiabatic exponent at a mass coordinate $m$ as:
\begin {equation}
\label{equ:gamma1}
\langle \Gamma_{1} (m) \rangle = \frac{\int_{m}^{M_{*}} \Gamma_{1} P dV} {\int_{m}^{M_{*}} PdV},
\end {equation}
where the first adiabatic exponent $\Gamma_{1}$ is defined as
\begin {equation}
\label{equ:gamma1def}
\Gamma_{1} = \left( \frac{\partial \mathrm{ln}\:P}{\partial \mathrm{ln}\:\rho}\right)_{S},
\end {equation}
and $P$ is pressure, $\rho$ is density, $S$ is entropy, $m$ is enclosed mass, $M_{*}$ is total star mass, and $V$ is volume.

The value of $\langle \Gamma_{1} \rangle$ is a measure of the dynamical stability, or instability, of a stellar model, with values of $\langle \Gamma_{1} (0) \rangle$ (that is, averaged over the entire star) below $^{4}\!/_{3}$ indicating models that are formally dynamically unstable against ejection or collapse \citep[see][]{Ritter1879, Ledoux1945, Stothers1999}. It has also been argued, although not formally proven, that $\langle \Gamma_{1} (m) \rangle$ for values of $m$ within the stellar envelope are indicative of instability to ejection from that point outward \citep[see][]{Lobel2001}. It is noteworthy that our models retain a high degree of dynamical instability throughout the pulsation cycle, with regions above the hydrogen and first helium ionisation zones keeping $\langle \Gamma_{1} (m) \rangle < \, ^{4}\!/_{3}$ in all phases except in the immediate aftermath of a cooling catastrophe.

The ejection seen in Figs~\ref{fig:fig10}--\ref{fig:fig12} is typical in that it is launched upon the surface breakout of a compression shock, which can be seen at approximately 64.6~years and a logarithmic radius of 2.2. The expansion following this compression is sufficiently fast to raise a shell of $\sim~0.09\mathrm{M}_{\odot}$ onto an escape trajectory. Material below the ejected layer expands on bound ballistic trajectories and eventually returns to the star with a supersonic fallback shock. This material becomes optically thin as it expands, a measure of which is given in Fig.~\ref{fig:fig11} as the logarithm of one tenth of the product of local radius, opacity, and density ($\frac{1}{10}r\kappa\rho$). The loss of optical depth of ejected and almost ejected-matter is not well accommodated by our stellar model, as the treatment of heat transport we employ is only strictly applicable in optically thick regions, but as this occurs after the ejection has been launched, we do not expect it to affect our results significantly.

We can also gain insight into the energetics of the ejection from these figures: the compression shock -- which can be seen most clearly in the plot of velocity divergence -- leaves the envelope's hydrogen completely ionized, and produces a region of singly ionized helium near the surface. In the shock's aftermath, there is heavy deposition of recombination energy into the expanding material -- the helium's energy is deposited very quickly and then the hydrogen's follows as the expansion continues. As the compression shock moves towards the surface, the kinetic energy of the pulsation's infalling material is temporarily stored in ionization of the envelope, and then released to accelerate the expansion of the model's outer layers in a recombination-powered bounce. The total amount of recombination energy released during the acceleration of this ejection is approximately $10^{46} \: \mathrm{ergs}$ in half a year, the majority of which occurs at optical depths greater than $10^4$ as can be seen by comparing panels 1 and 6 of Fig.~\ref{fig:fig11}. This recombination-powered bounce is analogous to the process of ``shell-triggered'' ejection seen in the 3-d hydrodynamical simulations of \citet{Ivanova2016}.

\subsection{Non-ejecting simulations}
\label{sec:nonejections}
For comparison, similar plots for the first supersonic compression of the $2.5 \times 10^{45}$~$\mathrm{ergs \: yr}^{-1}$ uniform heating model are shown in Figs.~\ref{fig:fig13}, \ref{fig:fig14} \& \ref{fig:fig15}.

This compression exhibits a strong compression shock, but does not result in a shell ejection. The shock in this compression is long-lived, and can be seen between approximately 41.7 and 43.1 years. In contrast to the ejecting case shown in Figs~\ref{fig:fig10}--\ref{fig:fig12}, this model exhibits a large degree of decoherence between the inner and outer layers of the envelope, with the inner regions collapsing significantly earlier than those near the surface. This decoherence leads the shock to form very deep within the envelope and become choked by the infalling matter of the outer layers. The shock then has a much weaker effect when it finally reaches the surface, and there is no coordinated rebound of matter from all but the outermost layers of the envelope.

Another major difference that can be seen in Fig.~\ref{fig:fig13} is that the infalling matter is almost completely neutral before it hits the shock, down to a mass coordinate of approximately 0.5~$\mathrm{M}_\odot$ -- for comparison, the neutral layer seen in Fig.~\ref{fig:fig10} reaches down only to mass coordinate of $\sim$1.45~$\mathrm{M}_\odot$. The very thick neutral layer seen in the non-ejecting case is a result of the large radiative energy losses sustained by that model when its radius was large during the cooling catastrophe. The kinetic and gravitational energy thermalized by the shock is used to reheat and reionize this neutral material, which has the effect of damping out the primary pulsation, as the energy stored in that pulsation mode is used for this purpose; in effect, the gravitational and kinetic energy released by the envelope's collapse is used to return the cool, neutral material to a (near-)thermal equilibrium state, after which additional energy is not available to drive a reexpansion. A comparison of the total envelope energies and ionization energies in our example non-ejecting and ejecting simulations can be seen in the bottom plots of Figs~\ref{fig:fig6} \& \ref{fig:fig7}, in which the total energy of the envelope (in purple), and the ionization energy of the envelope (in green) drop almost to zero during the cooling catastrophe in the non-ejecting case (Fig.~\ref{fig:fig6}), but remain much higher in the ejecting case (Fig.~\ref{fig:fig7}).

When the compression shock in this simulation reaches the surface, it does not display the helium ionization feature seen in Fig.~\ref{fig:fig10}, is not associated with a coherent rebound by the envelope's inner layers, and results in only a small reexpansion of the upper layers, which excites a short-lived higher-frequency pulsation that then decays.

Loss of coherence between the pulsations of different layers within the envelope and the loss of large amounts of energy during a cooling catastrophe are the main physical differences seen in pulsations which lack the strong rebound required to eject a mass shell. Our interpretation is that these effects act to damp the primary pulsation and prevent mass ejections in cases when the pulsation amplitude is high but ejections are not seen. Both effects occur whilst the star is expanded to very large radii (phases II and III), and are a result of the increase in pulsation period with radius -- the star's outer layers feel this effect more strongly than layers at lower radii, leading to decoherence, and a longer period means more time spent near maximum radius where radiative losses are strongest and the envelope can undergo catastrophic cooling.

It is worth noting that ejections also appear to be suppressed by the presence of incoherent pulsation modes within the envelope excited by shocks generated at previous pulsation minima. Thus the question of whether a given compression will launch a shell ejection depends heavily on the simulation's recent history. This is one way in which our simulations exhibit elements of chaos, with the future behaviour of a particular model highly sensitive to changes in initial conditions.

\begin{figure*}
	\includegraphics[width=1.5\columnwidth]{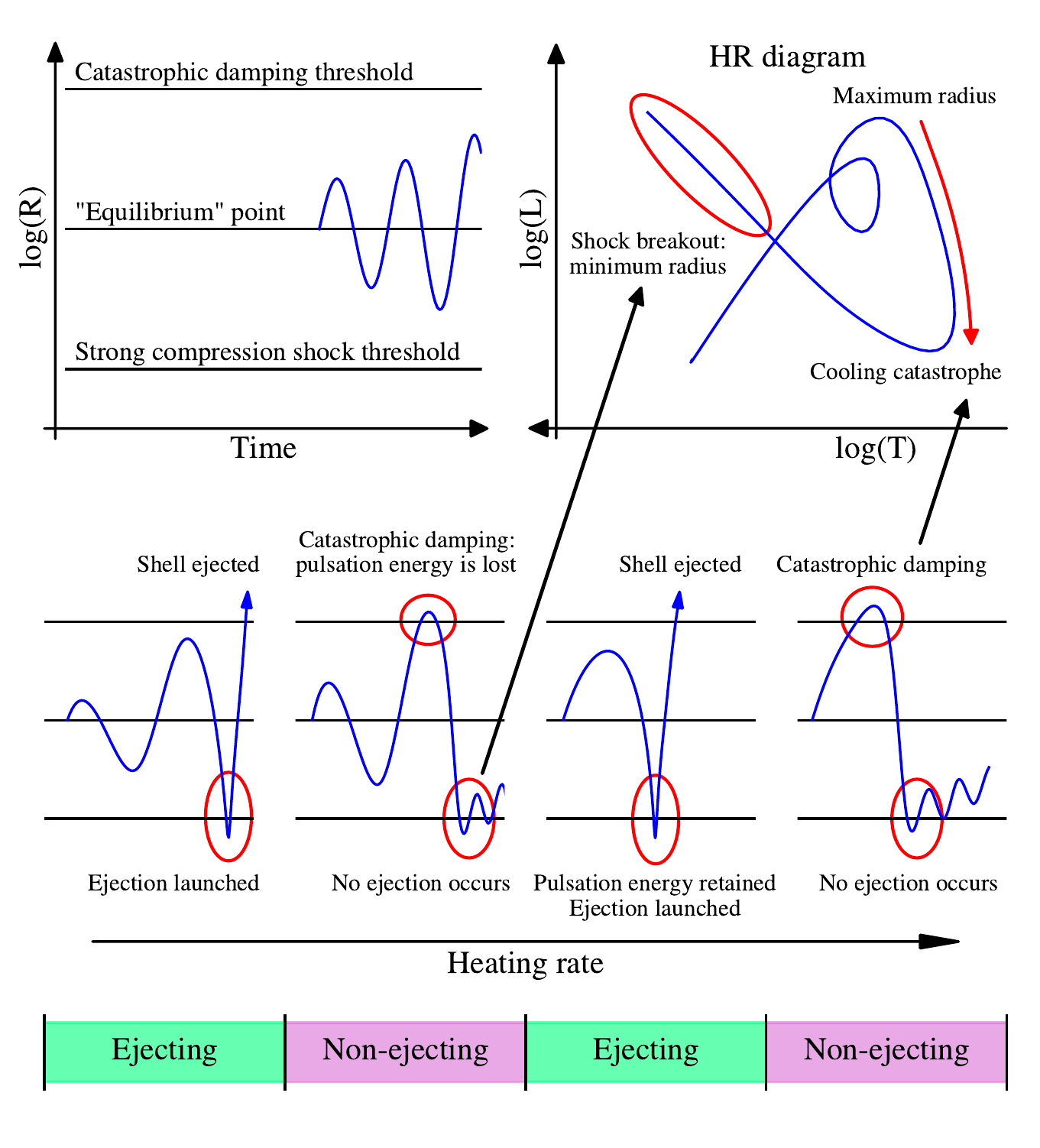}
    \caption{A schematic figure showing a simplified model of phenomena seen in the large-amplitude pulsations arising in our simulations. Four example evolutions of radius with time are plotted, representing four different heating rates which increase to the right. Upper and lower radius thresholds are shown which represent radii that must be achieved for a model to undergo catastrophic cooling and mass ejections respectively. In order for a star to eject mass, it must reach the lower threshold before the upper one. This gives rise to a complex structure of ejecting and non-ejecting regions in the heating rate parameter space. The major features of the simplified pulsation histories appear also on a simplified HR diagram showing the major features seen in HR diagrams of our simulation results, such as those shown in Fig.~\ref{fig:fig5}.}
    \label{fig:fig8}
\end{figure*}

\subsection{To eject or not to eject}
\label{sec:tobe}

\begin{figure}
	\includegraphics[width=1.05\columnwidth]{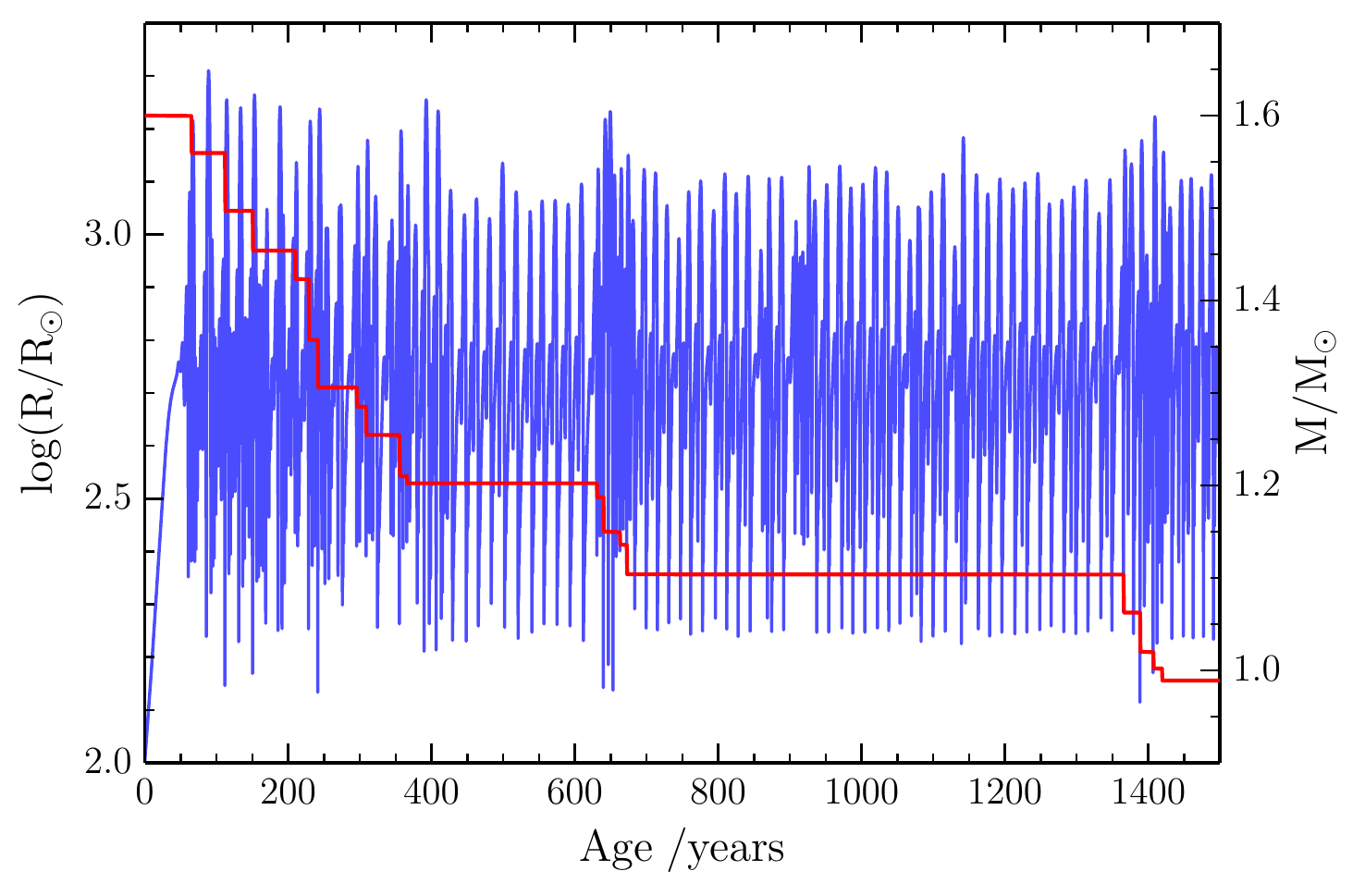}
    \caption{The surface radius (blue) and mass (red) of an extended simulation with uniform heating at $1.69 \times 10^{45}$~$\mathrm{ergs \: yr}^{-1}$, showing 18 successive ejections that together remove $\sim 51$\% of the initial mass of the envelope.}
    \label{fig:fig9}
\end{figure}

As we can see from Fig.~\ref{fig:fig3}, whether or not dynamical ejections are exhibited by a particular model is not a simple function of heating rate, and there is no single contiguous set of heating rates which encompasses all ejecting models (note for example that the $2.5 \times 10^{45}$~$\mathrm{ergs \: yr}^{-1}$ case does not exhibit ejections, despite being surrounded by models that do). The primary effect on a simulation from varying the heating rate is a change in the growth rate of pulsations; however, this growth is monotonically increasing with heating rate, so a more complex dependence of ejection on growth rate is required to explain these phenomena. In addition, some heating rates lead to multiple sequential ejections, as in the case of uniform heating at $1.7 \times 10^{45}$~$\mathrm{ergs \: yr}^{-1}$, whereas some lead to only one ejection followed by an apparently stable cycle of pulsation growth and shock dissipation. This behaviour would be expected to arise if the set of heating rates which leads to ejections changes with envelope mass, as seems intuitive, as these models change mass without changing heating rate.

Mass shell ejections occur when models rebound after compressions of sufficiently high amplitude (that is, compressions which are sufficiently deep), but are seen to be suppressed when these compressions display a high degree of decoherence between layers within the envelope, and when the internal energy of the envelope is very low. These two supressing phenomena both arise as a result of non-linear effects that emerge at very large pulsation amplitudes: very large amplitudes during the expanded phase of a pulsation lead to the decoherence of layers within the envelope, and to the radiative loss of large amounts of energy during a cooling catastrophe, and both of these effects act to prevent a strong rebound and to damp the primary pulsation. For a mass-ejection to occur, we require that the pulsation avoid being damped by these effects, but also that it have a sufficiently large amplitude at the minimum radius point in the pulsation, where the ejection is actually launched. In short, we require a pulsation whose amplitude is large, but not too large.

It is also important which phase the pulsation is in when it achieves maximum amplitude. The non-linear damping mechanisms described above both depend on the pulsation amplitude during the \emph{expanded} phase of the pulsation (phases II and III), whereas the ejection itself is launched in the trough of the compression (between phases III and IV). In order for ejections to occur, therefore, we require the pulsation amplitude to not be too high during the expansion phase, but also that it \emph{is} sufficiently high during the following compression phase. Because pulsations grow until they are damped by the non-linear effects discussed above, whether or not a given model will display mass ejections therefore depends on the interaction between the pulsation growth time-scale and the period of pulsation, as it is the interplay between these two time-scales that determines which phase the pulsation will be in when it reaches its amplitude peak. This means that the ranges of heating rates that will lead to ejections for a given model do not necessarily occupy a single contiguous region, but instead form a more complex shape, in a manner which can be thought of as forming resonances between the pulsation period and the amplitude growth time-scale.

A simplified ``toy'' model of how such a shape can emerge is described in Fig.~\ref{fig:fig8}. In this model, there are two radius thresholds, one high (outer) and one low (inner).

A star begins at an ``equilibrium'' radius between these two thresholds and pulsates around this initial point. The amplitude of this pulsation grows, and the star's behaviour depends on which thresholds it reaches, and at what times: a model which passes above the outer radius threshold will undergo sufficient catastrophic cooling and decoherence of its internal layers to damp out the primary pulsation, whereas a model which passes below the inner radius threshold has a sufficiently strong compression and associated shock to launch a shell-ejecting rebound, unless the pulsation has already been damped out. Therefore, in order for a shell to be ejected, a model must pass below the inner threshold \textbf{without first passing above the outer one}. As can be seen in this figure, such a model naturally gives rise to a complex structure of ejecting and non-ejecting regions in the heating-rate parameter space.

Models with certain heating rates exhibit multiple successive ejections, whilst others eject only once. A change in the mass of the model due to an ejection will likely cause a corresponding change in pulsation period, and possibly also the amplitude growth time-scale. Since the behaviour of the model depends on the interplay between these two time-scales, such changes are likely to cause the ranges of heating rates which produce ejections to shift, so the suppression of subsequent ejections in some models is unsurprising; the heating rate is held constant, but the regions of the parameter space may have moved such that that heating rate is now in a non-ejecting region.

An extended simulation of a heating case which exhibits repeated mass ejections is shown in Fig~\ref{fig:fig9}, which was run for 1,500 years of star time. This simulation experiences two extended periods with no ejection events, lasting first 265 and second 692 years. Although the short-term behaviour of the pulsations we report is chaotic, and therefore unpredictable, the presence of long periods with no ejection events, followed by the later reappearance of repeated mass ejections, suggests the possibility of an underlying structural relaxation occurring within the envelope during these quiescent periods, operating on a much longer time-scale than the pulsations of the model. However, a simple ``by eye'' analysis was unable to discern any significant changes in internal structure (e.g. in entropy, density, ionization profiles)  on top of the variation inherent in the pulsation. It should also be noted that the presence of long gaps between ejection events makes it difficult to say with certainty whether a given simulation will experience ejections in the future without calculating its evolution for an extended period of time.

\section{Discussion}
\label{sec:discussion}

\subsection{Numerical considerations}

We have not attempted to define the boundaries of the ranges of heating rates which will cause dynamical mass ejections. This is because the amplitude growth rate of pulsations is sensitive to changes in the numerical parameters of our simulations. We have performed convergence testing with regard to the temporal and spatial resolution of our models, the artificial-viscosity-induced shock thickness parameter ($l_2$), the model's outer boundary conditions, and the time-scale of mass-removal for ejected shells; although such changes do not alter the qualitative nature of our results, our simulations exhibit chaotic behaviour, and whether or not a given pulsation will exhibit mass ejections is highly sensitive to the initial conditions and simulation parameters of the simulation. Thus the regions of heating rate parameter space which lead to ejections, and the specific locations of the boundaries of these regions, can shift as a result of changes to the simulation parameters, and attempts to define the locations of these regions are not likely to be reliable (and this is before considering uncertainties introduced by our choice of model, see Section~\ref{sec:limitations}).

\subsection{Dynamical instability and mass loss}

Although our hydrostatic results match those of I15 closely, our hydrodynamic models evolve very differently in most cases. We see the same behaviour for the lowest heating rate of $10^{45} $~$\mathrm{ergs \; yr}^{-1}$: expansion followed by pulsational stability, and for the highest, $10^{47} $~$\mathrm{ergs \; yr}^{-1}$: direct dynamical ejection of the entire envelope. However, our models for the intermediate heating powers see neither the direct ejection of the base-heated hydrostatic case, nor the temporary stability of the uniformly-heated hydrostatic case. Instead, we find that there is a large region of parameter space with insufficient heating to drive a prompt ejection that produces models which are dynamically unstable to large-amplitude pulsations.

The pulsations we recover are similar to those obtained by authors such as \citet{Wood1974} and \citet{Tuchman1978, Tuchman1979} in the context of dynamically unstable single giants, which likewise display repeated mass-loss events in the form of dynamical shell ejections. By applying the shock treatment model those works lacked, our results represent a corroboration their findings with modern simulation techniques. More recent work published by \citet{Heger1997} and \citet{Yoon2010} has reported the emergence of similar pulsations growing in models of red supergiants (RSG), but those authors terminated their simulations before they could be affected by shocks, so could not study the shock-dominated regimes reported in this work.

The shells ejected by our models have masses of up to $\sim$~$0.1 \mathrm{M}_{\odot}$. In some cases these ejections are repeated within a few decades, leading to an effective time-averaged mass-loss rate of order $10^{-3}$~$\mathrm{M}_{\odot} \: \mathrm{yr}^{-1}$. This is sufficiently high to clear the entire envelope of our model star within approximately 1000 years, making the shell ejections seen here a possible mechanism for the delayed ejection of common envelopes during the duration of the slow spiral-in phase. Although some of our models exhibit single, rather than repeated ejections, we believe the suppression of further ejections in these cases to be an effect of changing the envelope's mass, and therefore its pulsation characteristics, putting the heating rates of those models into non-ejecting regions of the parameter space. In reality, we would not expect heating rates to remain constant when such ejections occur, as in our models, so the problem of stars becoming ``stuck'' at non-ejecting heating rates is not likely to be relevant.

\subsection{Ejection efficiency}
The situation we model in this work represents a highly non-adiabatic phase of the CE process, and is therefore not the situation for which the $\alpha$ efficiency parameter was originally conceived. However, it is still possible to define an equivalent efficiency value for the mass-ejection process that emerges from our calculations.

One common definition of the efficiency parameter is

\begin{equation}
\label{equ:alphadef}
\alpha \, \Delta E_\mathrm{orb} = E_\mathrm{bind} \approx \frac{G M_\star M_\mathrm{env}} {\lambda R_\star} ,
\end{equation}
where $\Delta E_\mathrm{orb}$ is the energy lost from the orbit during the CE process, $E_\mathrm{bind}$ is the gravitational binding energy of the envelope (for the purposes of this calculation, this does not include contributions from thermal energy), $G$ is the gravitational constant, $M_\star$ and $R_\star$ are the mass and radius of the giant, $M_\mathrm{env}$ is the mass of the giant's envelope, and $\lambda$ is a parameter commonly used to encode the dependence of the envelope's gravitational binding energy on the internal structure of the giant \citep[see][]{DeKool1990}, equivalent to $\lambda_\mathrm{g}$ in \citet{Dewi2000}.

In this way, a proportion $\alpha$ of the energy lost from the binary orbit is equal to the energy required to eject the envelope. Problems in calibrating a values for $\alpha$ and $\lambda$, and in determining the correct value of $E_\mathrm{bind}$ (which depends on, for example, the precise location of the envelope/core transition, the amount of kinetic energy left in the ejected material, the amount of energy that is radiated away from the CE object and the internal energy sources available within the envelope), render this formalism inaccurate for making predictions, but it remains useful for its simplicity and ease of application.

A value of the efficiency parameter for the mass-ejection process seen in our simulations can be defined as the ratio of the binding energy that was possessed by the ejected material prior to an ejection event and the heating energy put into the system in the run-up to a that event.

An approximate expression for this might be

\begin{equation}
\alpha_\mathrm{eject} = - \frac { m_\mathrm{eject} \epsilon_\mathrm{grav} } { P_\mathrm{heating} \tau_\mathrm{eject} } ,
\end{equation}

\noindent where $m_\mathrm{eject}$ is the mass of the shell that becomes unbound in an ejection event, $\epsilon_\mathrm{grav}$ is the mass-averaged specific gravitational energy of the envelope, $P_\mathrm{heating}$ is the rate at which the envelope receives synthetic orbital heating, and $\tau_\mathrm{eject}$ is the time between ejection events. In this case, if the entire envelope is removed by repeated shell ejections, we would have

\begin{equation}
\alpha \approx \frac{1}{M_\mathrm{env}} \sum \alpha_\mathrm{eject} m_\mathrm{eject} .
\end{equation}

It is not meaningful to take the average of the specific gravitational energy of the envelope at any specific point during the pulsation, as it varies a lot over any one cycle, so we will calculate it first for the initial RG model used as the starting point of all our simulations, and then for an ``equilibrium'' model, after the envelope has expanded to approximately the radius about which pulsations oscillate.

Using an extended simulation with a heating rate of $1.7 \times 10^{45}$~$\mathrm{ergs \: yr}^{-1}$ and uniform heating, covering 500 years and averaging over 14 consecutive ejections which together removed $\approx 0.6 \,\mathrm{M}_{\odot}$ of envelope material, we calculated the average efficiency of the ejection process. We find an $\alpha$ value of 0.25 and a $\lambda$ of 0.55 when gravitational binding energy is calculated relative to the initial RG model, and an $\alpha$ of 0.046 and $\lambda$ of 0.11 when calculated relative to an expanded radius of 518 $\mathrm{R_\odot}$ at 40 years. For purposes of comparison to existing literature, the first of these two definitions (comparison to the initial RG model) is closer to the one most often used by other authors.

To get an idea of how these values can vary, we performed the same calculation using a 500 year simulation with a heating rate of $3 \times 10^{45}$~$\mathrm{ergs \: yr}^{-1}$ and uniform heating, the greatest heating rate for which successive ejections were seen. This simulation yields an average $\alpha$ value of 0.046 and a $\lambda$ of 0.55 when calculating energies relative to the initial RG model, and an $\alpha$ of 0.013 and $\lambda$ of 0.075 when calculated relative to an expanded radius of 740 $\mathrm{R_\odot}$ at 20 years. This simulation ejected a total of mass of approximately $0.27 \,\mathrm{M}$ over 6 successive ejections. It should be noted that, in addition to the efficiency values varying between different simulations, the values calculated for each specific ejection \emph{within} a given simulation also vary considerably due to variation in the time (and therefore energy input) between ejections, and the masses of the ejected shells.

The lower energy efficiency in this second simulation is a result of there being a larger average time between ejections, as well as a greater energy input rate, both of which increase the total energy input per ejection. It is noteworthy that a decrease in both average mass-loss rate and ejection efficiency should be a result of \emph{increasing} the rate at which energy is injected into the model, which does not at first appear intuitive. The reduced ejection efficiency seen here is indicative of increased loss of energy to radiation, which is accelerated at higher heating rates by greater expansion of the envelope and higher average surface luminosities.

Both of these estimates for the ejection efficiency are considerably lower than the values commonly used to model the CE process in BPS codes, which tend to assume the product $\alpha \lambda$ be $\approx 1$ \citep[e.g.][]{Kinugawa2014, Hurley2002, Belczynski2008}. However, the values described in this section are specific to the slow spiral-in phase and therefore relate only to CE systems that enter this phase, and not to systems that experience envelope ejection during the fast plunge-in phase. The different ejection mechanisms that are likely to apply in these two phases make it probable that different ejection efficiencies are to be expected. A simple CE prescription for BPS codes which takes this difference into account can be imagined: one value of the efficiency parameter could be applied for CE systems that are expected to undergo envelope ejection during the fast plunge-in phase, while a second value of the efficiency parameter could be applied instead to systems expected to enter the slow spiral-in phase, utilising some (as yet unknown) criterion for determining which category a given system falls into.

\subsection{Limitations of our simulations}
\label{sec:limitations}

The model we have employed in this work depends on several major simplifications of the CE process. Primarily, our simulations are confined to one dimension. The main problems caused by this simplification of geometry affect the modelling of the secondary's orbit, which we make no attempt to follow in this work, but we would also expect the presence of the secondary at one particular position in the envelope to have additional effects on the envelope in the forms of rotation, gravitation and localised heating, none of which we include in our simulations. \citet{Ivanova2016} found that the envelopes of 3-d CE models which entered the slow spiral-in phase exhibited non-negligible differences in density between polar and equatorial directions, but that these differences were considerably smaller than the up to order of magnitude differences seen during the plunge-in phase, which we are used to seeing in 3-d simulations. Despite the inability of our model to treat these asymmetries, it is likely to exhibit the correct qualitative behaviour, with quantitative differences from reality due to geometric effects. It is possible that the complete 3-d system may actually be more susceptible to the kind of mass ejections we find, as its rotation will reduce the energy requirements for mass loss in the equatorial plane. In addition, it is important to remember that the frictional heating, which is averaged across spherical mass shells in our models, is expected to be more localised in the true 3-d system; our results have proven largely insensitive to the radial location of heating within the envelope, but it is possible that localisation of this heating in 3-d will have a stronger effect on the pulsational properties of the envelope.

An important feature missing from our models is that frictional heating during the slow spiral-in phase is not expected to be constant, but to depend on the density of material at the radius of the secondary's orbit. The large-amplitude pulsations we recover are therefore likely to lead to considerable variation in heating rate. However, as the true heating rate is expected to increase with density, it should be higher during phases of envelope compression; this effect therefore represents a powerful potential \emph{driving} mechanism for pulsations, and is likely to enhance the effects we see in our simulations. It is also worth noting that our simulations are based on heating zones defined at constant Lagrangian mass coordinates within the envelope, whereas in reality the region in which heating occurs should be better defined with radial coordinates, as it is dependent on the position of the secondary's orbit.

Another major consideration is the implementation of convection in our models, which is not properly time-dependent. It is ``time-dependant'' in a bad way, as it responds to changes in convective stability by turning convection on and off instantaneously, but this is clearly not an accurate representation of the physics. The lack of a true time-dependent implementation of convection is a major source of uncertainty in our results, as differences in energy transport are liable to have a significant effect on the growth rates of pulsations; however, convective turnover times in our models are generally a fraction of a year, a time-scale two orders of magnitude below the long-period pulsations we observe. Although there exist candidate treatments for time-dependent convection applicable to stellar evolution codes \citep[see, for example][]{Grigahcene2005}, these are not yet ready for full deployment \citep[see][]{Gastine2011}. In a study of analogous long-period pulsations in models of red supergiants, \cite{Heger1997} found that both pulsation periods and growth rates were largely unaffected by whether convective fluxes adjusted instantly to the stellar structure or were completely frozen in. Furthermore, it has been shown that in linear stability analysis calculations, pulsation properties do not vary strongly when an artificial lag between pulsation phase and convective flux is added \citep{Langer1971}.

\subsection{Dynamical shell ejections in other classes of stars}

The presence of dynamical mass ejections in our simulations raises the question of whether they would occur in models of other unstable giant stars.

Asymptotic giant branch (AGB) stars are a particularly interesting example, as we know that many of them exhibit pulsational instability on time-scales of several hundred days (e.g. Mira variables or long-period variables). Indeed it seems likely that the pulsations in our models arise from the same instability as in these stars, and for this reason it was in the context of AGB stars that much of the pioneering work on simulating unstable giants occurred \citep[e.g.][]{Wood1974, Tuchman1979}. Although it is believed today that the majority of mass loss on the AGB occurs on much longer time-scales and is due to wind acceleration in the extended atmospheres of these stars \citep[for a summary of this development, see][]{Kwok2011}, recent analyses of planetary nebulae have noted the existence of ``rings'' or ``arcs'', which are indicative of several episodes of strong, highly periodic mass loss from the progenitor AGB stars at the end of that phase, leading to discrete shells of matter in (proto-)planetary nebulae \citep[see, for example][]{Corradi2004, Phillips2009}. Future work may be able to determine if dynamical mass ejections can provide a good model for this phenomenon. Another attractive feature of these stars is that the nuclear luminosity variation produced by the thermal pulse cycle moves them through large regions of luminosity parameter space, making it more likely that they are susceptible to dynamical instability for at least part of that cycle.

Another possible candidate for these kind of pulsations are RSG stars. \citet{Yoon2010} have recently found similar-period pulsational instabilities in hydrodynamical models of these stars, and analyses of SNe such as SN 1979C \citep{Weiler1992} have revealed variable mass-loss prior to core collapse. Pulsation-induced mass loss is a strong candidate model for explaining such phenomena, possibly through mechanisms such as case D mass-transfer to a binary companion \citep[see][]{Mohamed2007}.

It is also possible that a similar process has occurred in the enigmatic luminous blue variable system $\eta$ Carinae. The multiple epochs of mass ejection reported by \citet{Kiminki2016} require that the mechanism causing this system's outbursts reoccur multiple times over at least 600 years. Our results, in particular that an episodic series of mass ejections can be generated by a constant rate of heating, suggest that it may be possible to explain the behaviour of this system using a similar model to the one presented here.

\subsection{Observational signatures}
It is possible to make some predictions about what the objects modelled in this work would look like if observed. Most notably they would exhibit strong variability over periods in the range 3--20 years. It is possible that this variability would be complex and quasi-periodic, especially for larger amplitudes. Due to this complexity, and the large variation seen between pulsations at different heating rates, it is not possible to make detailed predictions about the light-curve shapes of these pulsations. However, it is valuable to note that the pulsations we report are highly asymmetric; in particular, the high-luminosity sections of the pulsations persist for much longer than the low-luminosity sections, especially at large amplitudes.

Using the surface properties of our models we can estimate that the central points of these objects' pulsations would occupy a region of the HR diagram with $\log T_\mathrm{eff}$ between 3.4 and 3.5 and $\log L$ between 4.0 and 4.4. The models which display the largest pulsation amplitudes (associated with large heating rates) have values of $\log L$ that vary between approximately 2.5 and 5.2, with lower amplitude, non-shock-dominated models with lower heating rates saturating at a variation between approximately 3.3 and 4.6. The values of $\log T_\mathrm{eff}$ of the highest amplitude pulsators vary between approximately 3.2 and 3.8, with non-shock dominated models occupying the region between 3.5 and 3.7.

These predictions, however, are based solely on the surface properties of the envelope. As our models eject shells of matter containing a significant fraction of their total mass, it is possible that the envelope's surface will be heavily obscured by this ejected material, which may have a large effect on the observable properties of these objects. In particular, as the ejected mass shells continue to expand and cool, we would expect them to form large amounts of dust, which may obscure the central CE object entirely in the visible spectrum, rendering it observationally similar to an OH/IR star. The beginning of this cooling process can be seen in Fig.~\ref{fig:fig7}, and an increase in opacity within the ejected shell (in this case due to molecule formation) can be seen within one year of the launch of the ejection in Fig.~\ref{fig:fig12}.

\section{Conclusions}
\label{sec:conclusions}

We have carried out stellar evolutionary calculations of a 1.6~$\mathrm{M}_{\odot}$ red giant undergoing a synthetic CE event. By applying additional heating to the star's envelope, we simulate the presence of a 0.3~$\mathrm{M}_{\odot}$ companion embedded within it during the slow spiral-in phase. We applied this heating at rates ranging from $10^{45}$--$10^{47}$~$\mathrm{ergs \: yr}^{-1}$, representing spiral-in times of 10--1000 years, and distributed it either throughout the convective envelope or in a thin shell at the envelope's base. The response of the giant was modelled using the stellar evolution code \texttt{MESA} in both the hydrostatic regime and then using a shock hydrodynamics treatment.

Our hydrostatic results match closely those obtained using an independent code by \cite{Ivanova2015}, whose initial model we reproduced. In these simulations, models heated at the base of the envelope tended to expand promptly to escape velocity, whilst those heated throughout the envelope expanded by a factor of $\sim 10$ in radius before destabilising. The highest and lowest heating rates examined behaved very similarly in both cases, expanding promptly to escape velocity or expanding into a non-pulsating equilibrium respectively. These results indicate the necessity of a hydrodynamical treatment for this scenario, as dynamical effects quickly become dominant in the models' evolution.

Our hydrodynamical simulations revealed that almost the entire range of heating rates we investigated causes the expanded giant envelope to become dynamically unstable to pulsations with periods in the range 3--20 years and extremely high growth rates, whose time-scales range from of order 10 pulsation periods to less than one period. These pulsations grow into a supersonic regime and develop strong compression shocks that pass outward through the outer layers of the envelope, and are eventually damped by the non-linear effects of catastrophic cooling and internal decoherence that emerge at large pulsation amplitudes. In certain cases, the rebound following a high-amplitude compression can be strong enough to accelerate a layer of matter at the star's surface to above escape velocity, dynamically ejecting a shell of matter from the star. This rebound is partially powered by the recombination of material ionized during the compression. The shells ejected in this way can be up to 0.1~$\mathrm{M}_{\odot}$ in mass, and ejections can repeat within a few decades, leading to a time-averaged mass-loss rate of order $10^{-3}$~$\mathrm{M}_{\odot} \: \mathrm{yr}^{-1}$. This mass-loss rate is sufficiently high to represent a candidate mechanism for removing the entire envelope over the duration of the slow-spiral in phase, and represents an $\alpha \lambda$ ejection efficiency in the range $0.025-0.14$, with $\alpha$ in the range  0.046-0.25.

Additional work is needed to extend the simulations performed here to longer time-scales and to assess the viability of repeated dynamical shell ejections for removing a majority of the mass of the envelope, as well as to increase the fidelity of the CE model, in particular by adding feedback between envelope structure and orbital heating rate, which is likely to have a strong effect on the growth of pulsations. The study of this phenomenon in models of unstable giants not undergoing CE events may also yield interesting results.

\section{Acknowledgements}

MC acknowleges the support of the Science \& Technology Facilities Council through grant ST/K502236/1. NI acknowledges that this research was undertaken, in part, thanks to funding from the Canada Research Chairs program, and funding from NSERC Discovery. SJ is partially supported by the Strategic Priority Research Program of the Chinese Academy of Sciences ``Multi-waveband Gravitational Wave Universe'' (Grant No. XDB23040000), and is also grateful to the Chinese Academy of Sciences (President's International Fellowship Initiative grant No.2011Y2JB07) and the National Natural Science Foundation of China (grant Nos. 11250110055, 11350110324, and 11633005) for funding. PhP, NI and SJ acknowledge that part of this work was performed at the Aspen Center for Physics, which is supported by National Science Foundation grant PHY-1066293. NI \& SJ gratefully acknowledge the hospitality of ISSI Bern during discussions of this work. This research was also partially undertaken at the Kavli Institute for Theoretical Physics which is supported in part by the National Science Foundation under Grant No. NSF PHY11-25915.




\bibliographystyle{mnras}
\bibliography{bibliography}


\begin{figure*}
	\includegraphics[width=2.15\columnwidth]{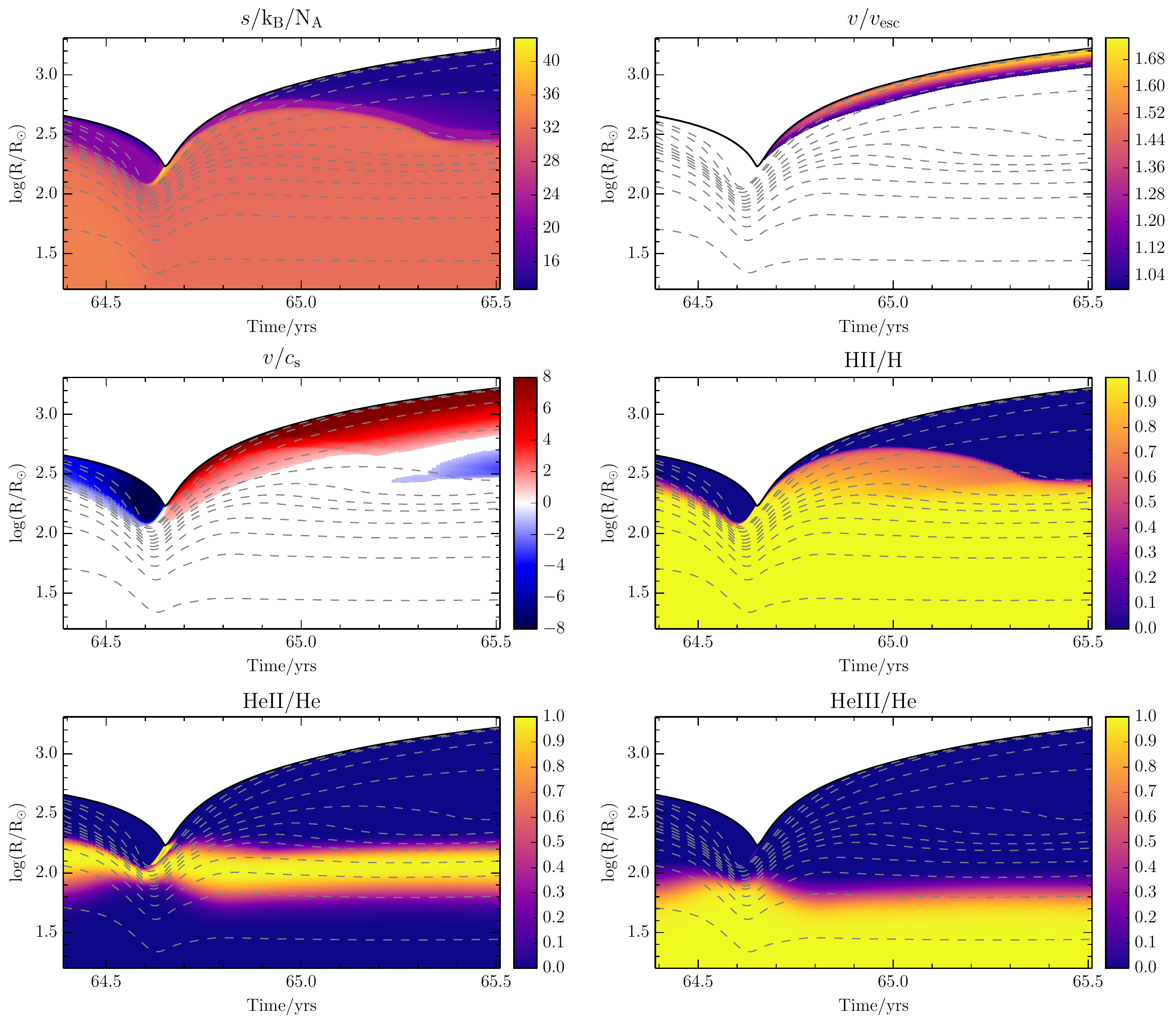}
    \caption{The first mass ejection displayed by a model heated uniformly throughout the convective envelope at $1.7\times10^{45}$~$\mathrm{ergs \: yr}^{-1}$ showing non-dimensionalised entropy per particle; the ratio of velocity and local escape velocity for regions where this ratio is above 1; the ratio of velocity and local sound speed for supersonic regions; and the relative proportions of ionised hydrogen and singly and doubly ionised helium. Also shown are contours containing 100\% in black, and 99, 98, 95, 90, 85, 80, 75, 70, 60, 50, 40, and 30\% in dashed grey, of the total mass of the model.}
    \label{fig:fig10}
\end{figure*}

\begin{figure*}
	\includegraphics[width=2.15\columnwidth]{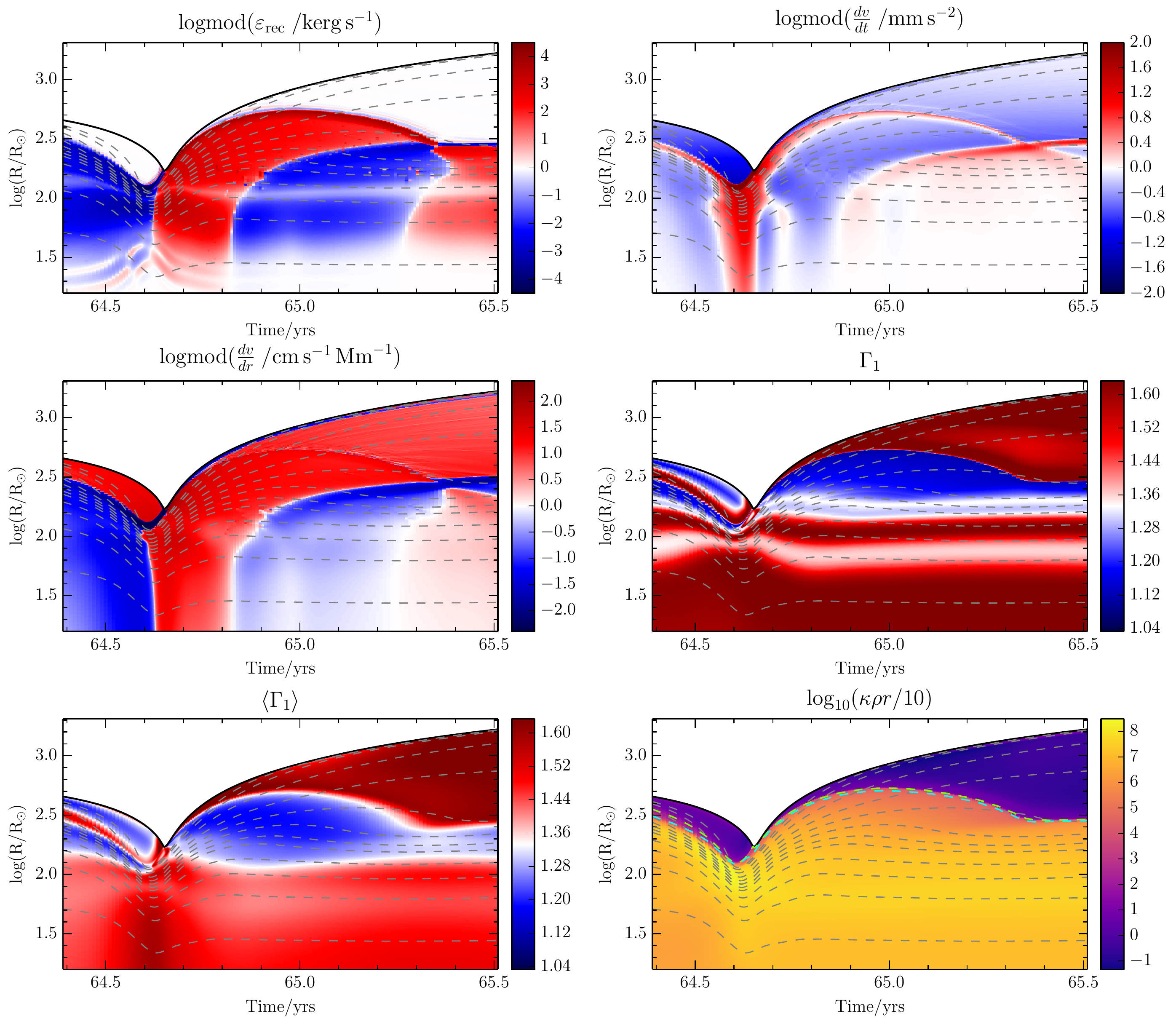}
    \caption{For the same event shown in Fig~\ref{fig:fig10} (the first mass ejection displayed by the model heated uniformly throughout the convective envelope at $1.7\times10^{45}$~$\mathrm{ergs \: yr}^{-1}$), we show the specific release rate of recombination energy; acceleration; velocity divergence; the first adiabatic exponent with white corresponding to the critical value of 4/3; the pressure-weighted, volume averaged value of the first adiabatic exponent (see Equation~\ref{equ:gamma1}) with white corresponding to the critical value of 4/3; and one tenth of the product of opacity, density and local radius, a dimensionless quantity representative of the local optical thickness of the stellar material. Panel 6 also shows the radii at which the optical depth of the star is $10^2$ in dashed green, and $10^4$ in dashed cyan. Also shown in all panels are contours containing 100\% in black, and 99, 98, 95, 90, 85, 80, 75, 70, 60, 50, 40, and 30\% in dashed grey, of the total mass of the model.}
    \label{fig:fig11}
\end{figure*}

\begin{figure*}
	\includegraphics[width=2.15\columnwidth]{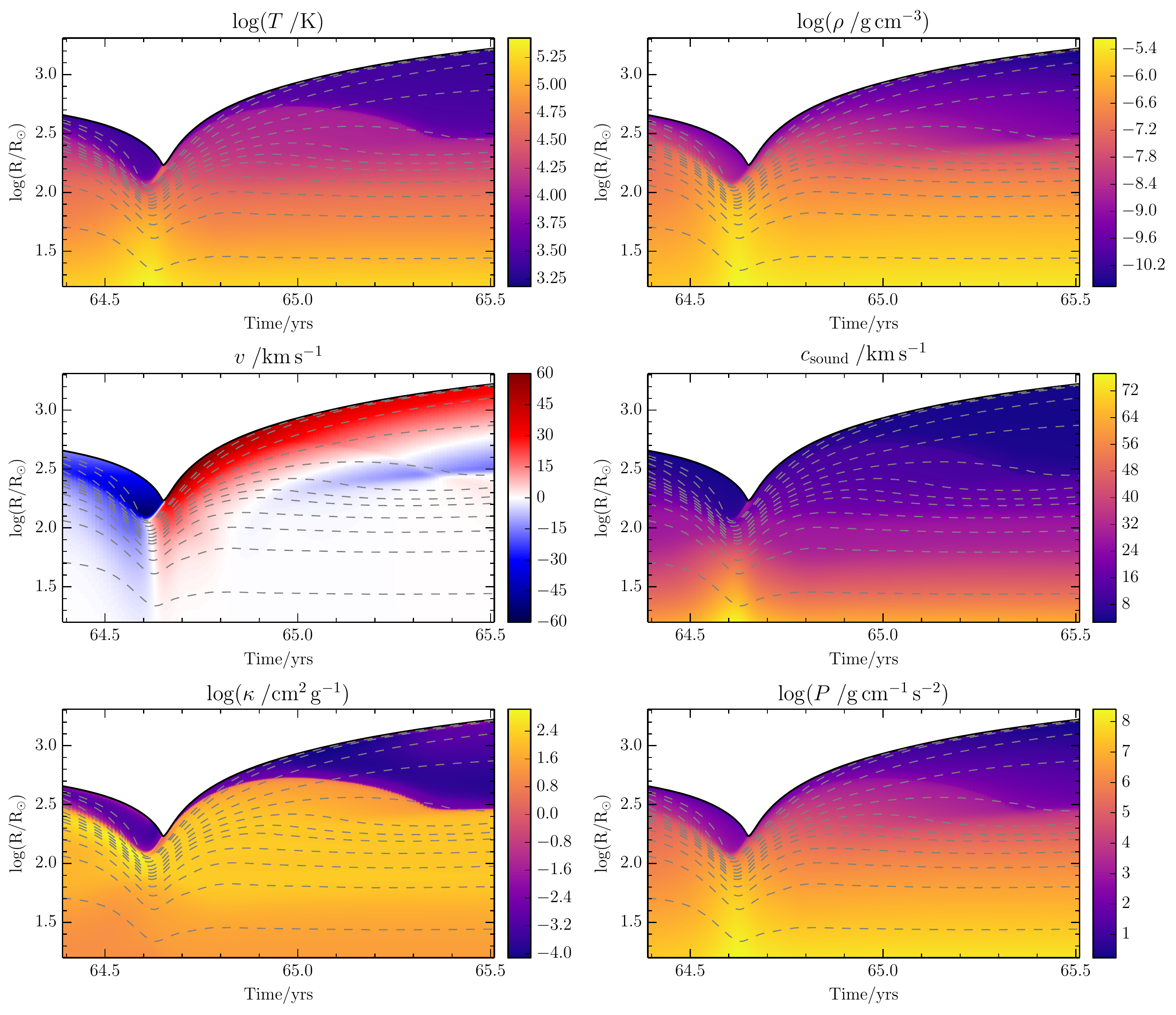}
    \caption{For the same event shown in Fig~\ref{fig:fig10} (the first mass ejection displayed by the model heated uniformly throughout the convective envelope at $1.7\times10^{45}$~$\mathrm{ergs \: yr}^{-1}$), we show the temperature; density; velocity; local sound speed; opacity; and pressure. Also shown in all panels are contours containing 100\% in black, and 99, 98, 95, 90, 85, 80, 75, 70, 60, 50, 40, and 30\% in dashed grey, of the total mass of the model.}
    \label{fig:fig12}
\end{figure*}

\begin{figure*}
	\includegraphics[width=2.15\columnwidth]{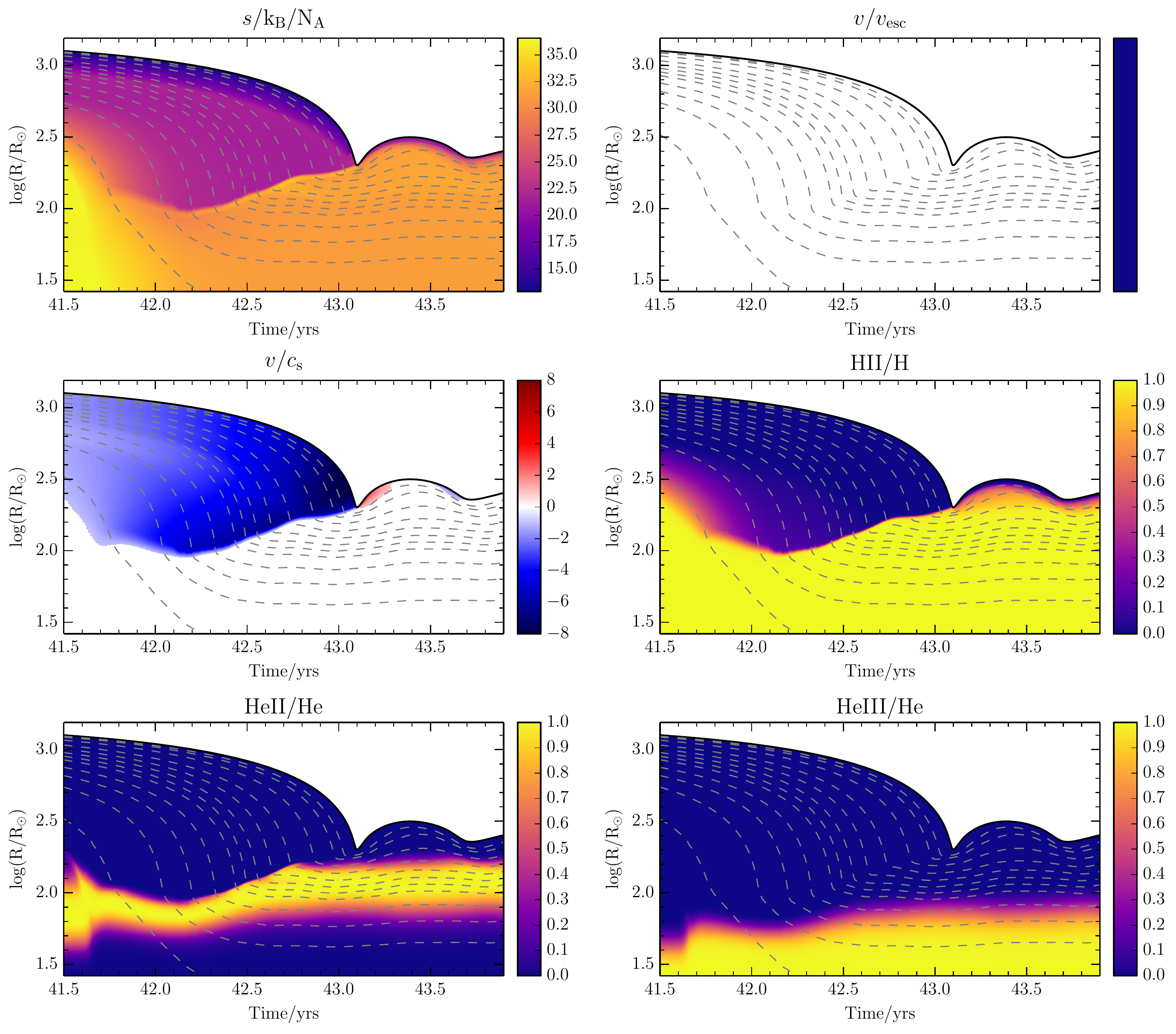}
    \caption{The first catastrophic damping and non-ejecting collapse episode experienced by a model heated uniformly throughout the convective envelope at $2.5\times10^{45}$~$\mathrm{ergs \: yr}^{-1}$, showing non-dimensionalised entropy per particle; the ratio of velocity and local escape velocity for regions where this ratio is above 1 (which occurs nowhere in this plot); the ratio of velocity and local sound speed for supersonic regions;  and the relative proportions of ionised hydrogen and singly and doubly ionised helium. Also shown are contours containing 100\% in black, and 99, 98, 95, 90, 85, 80, 75, 70, 60, 50, 40, and 30\% in dashed grey, of the total mass of the model.}
    \label{fig:fig13}
\end{figure*}

\begin{figure*}
	\includegraphics[width=2.15\columnwidth]{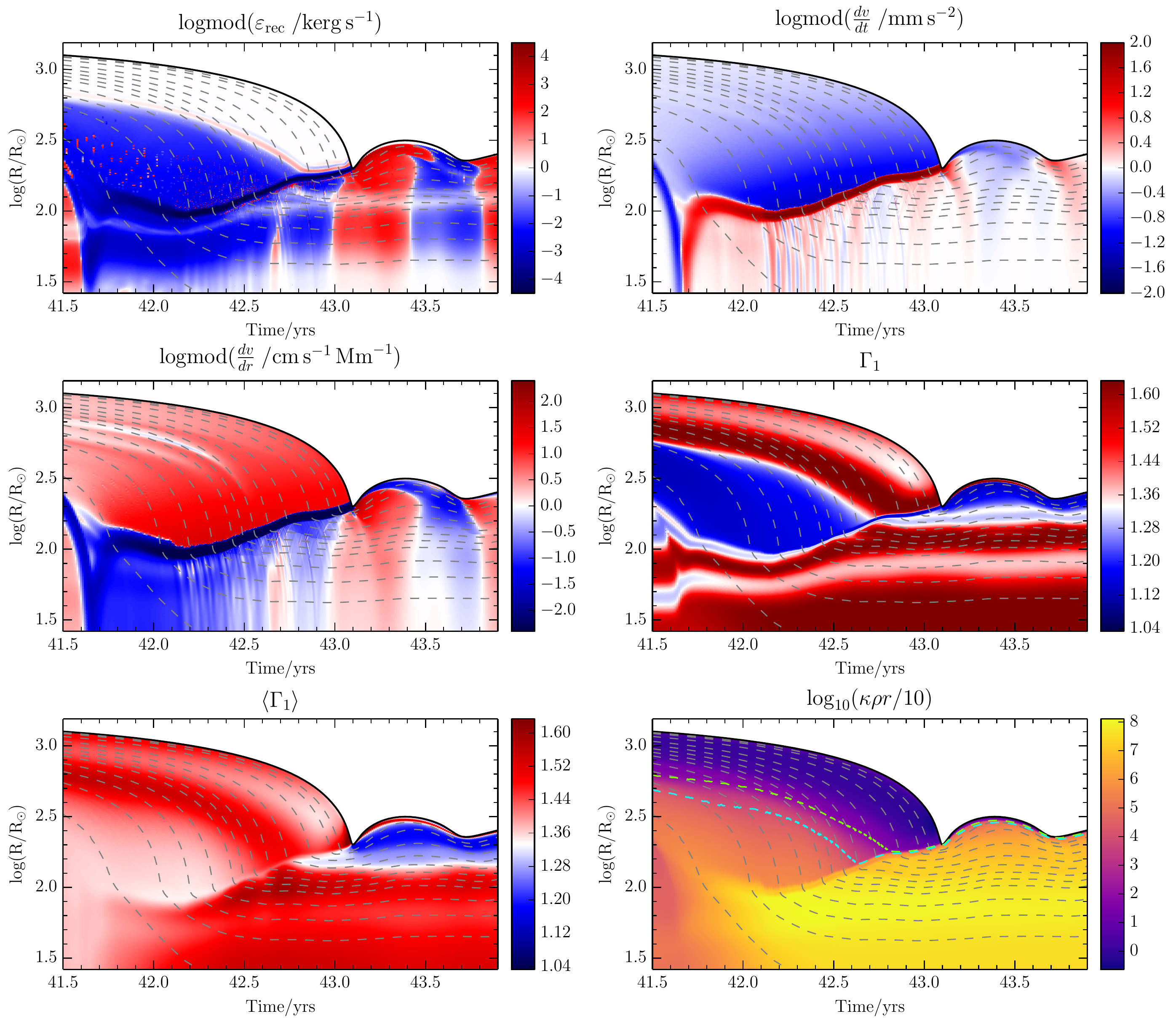}
    \caption{For the same event shown in Fig~\ref{fig:fig13} (the first catastrophic damping and non-ejecting collapse episode experienced by a model heated uniformly throughout the convective envelope at $2.5\times10^{45}$~$\mathrm{ergs \: yr}^{-1}$), we show the specific release rate of recombination energy; acceleration; velocity divergence; the first adiabatic exponent with white corresponding to the critical value of 4/3; the pressure-weighted, volume averaged value of the first adiabatic exponent (see Equation~\ref{equ:gamma1}) with white corresponding to the critical value of 4/3; and one tenth of the product of opacity, density and local radius, a dimensionless quantity representative of the local optical thickness of the stellar material. Panel 6 also shows the radii at which the optical depth of the star is $10^2$ in dashed green, and $10^4$ in dashed cyan. Also shown in all panels are contours containing 100\% in black, and 99, 98, 95, 90, 85, 80, 75, 70, 60, 50, 40, and 30\% in dashed grey, of the total mass of the model.}
    \label{fig:fig14}
\end{figure*}

\begin{figure*}
	\includegraphics[width=2.15\columnwidth]{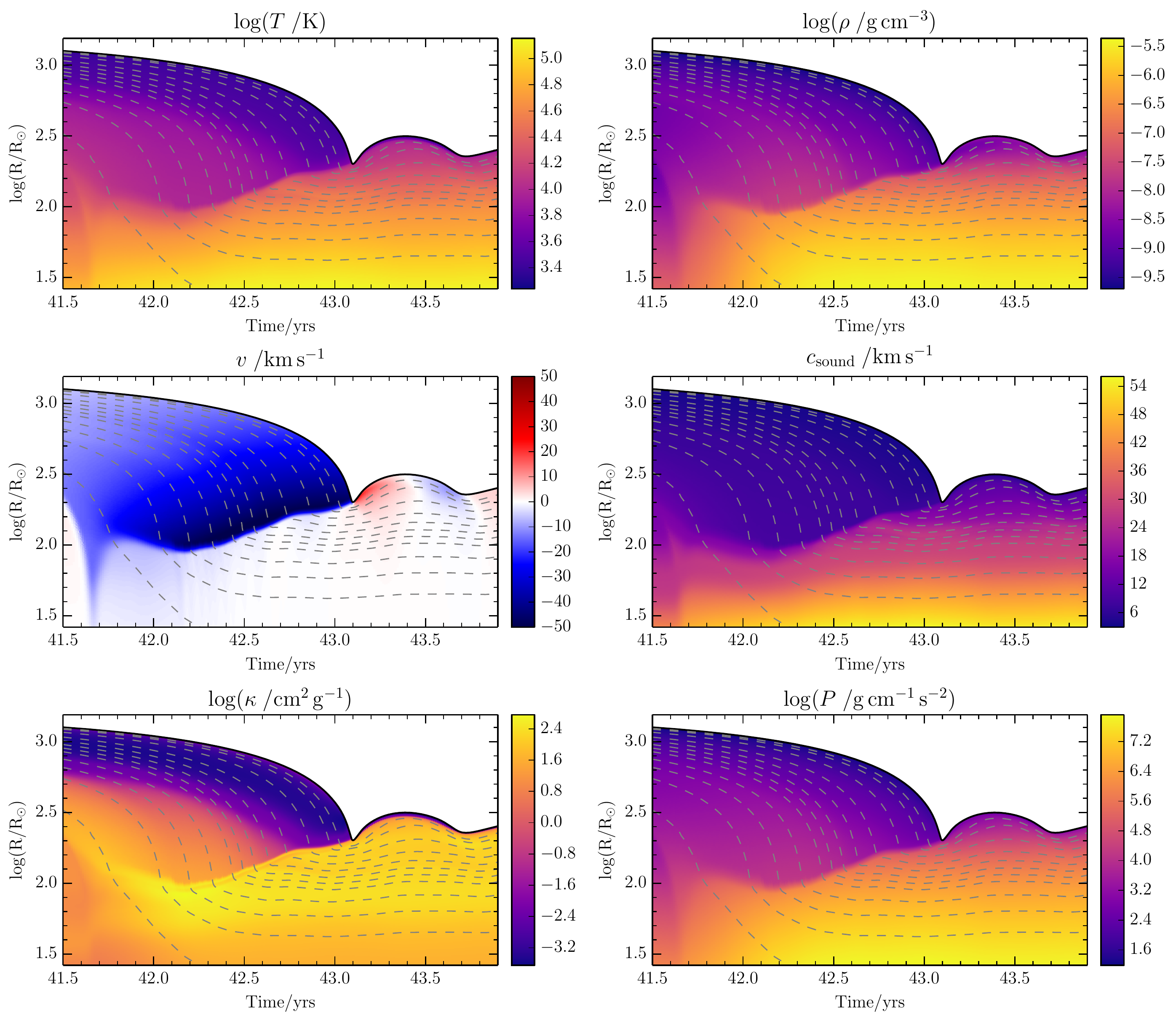}
    \caption{For the same event shown in Fig~\ref{fig:fig13} (the first catastrophic damping and non-ejecting collapse episode experienced by a model heated uniformly throughout the convective envelope at $2.5\times10^{45}$~$\mathrm{ergs \: yr}^{-1}$), we show the temperature; density; velocity; local sound speed; opacity; and pressure. Also shown in all panels are contours containing 100\% in black, and 99, 98, 95, 90, 85, 80, 75, 70, 60, 50, 40, and 30\% in dashed grey, of the total mass of the model.}
    \label{fig:fig15}
\end{figure*}


\appendix
\FloatBarrier
\newpage
\onecolumn
\section{Mass-loss routine}
\label{app:mdot}
The custom mass-loss scheme used in this work is implemented in MESA using the other\_wind hook routine, which is called at every timestep to find the appropriate amount of mass to remove from the star during that step. The algorithm used to calculate that mass is reproduced below using Fortran syntax.

\begin{lstlisting}
   ! m(x) is the mass coordinate of cell x
   ! r(x) is the radius coordinate of cell x
   ! v(x) is the velocity of cell x
   ! G is Newton's gravitational constant
   ! msol is one solar mass
   ! cellnum is the total number of cells in the model
   ! mdot is the mass-loss rate to be calculated
 
   limit_cell = 0
 
   ! Cells are numbered from 1 at the star's outer edge, to cellnum at the centre
   do x = 1, cellnum

      v_esc = sqrt( 2 * G * m(x) / r(x) )

      if ( v(x) > v_esc ) then
         limit_cell = x
      else
         exit
      end if

   end do

   if ( limit_cell == 0 ) then

      mdot = 0

   else

      mdot = ( m(1) - m(limit+1) ) / msol * 100   ! in solar masses per year

   endif

\end{lstlisting}


\bsp	
\label{lastpage}
\end{document}